\documentclass[aps,prd,twocolumn,showpacs,nofootinbib,floatfix,nolinenumbers]{revtex4-1}

\usepackage{atlasphysics}
\usepackage[hyperindex,breaklinks]{hyperref}
\usepackage{subfigure}
\usepackage{mathrsfs}
\usepackage{units}
\usepackage{graphics}
\usepackage{graphicx}
\usepackage{amsmath}
\usepackage{url}
\usepackage{xspace}

\newcommand{\gravitino}{\ensuremath{\tilde{G}}\xspace}

\newcommand{\Ref}{Ref.}

\newcommand{\zdca}{\ensuremath{\Delta z_\gamma}\xspace}
\newcommand{\zDCA}{\ensuremath{\Delta z_\gamma}\xspace}
\newcommand{\abszdca}{\ensuremath{|\Delta z_\gamma|}\xspace}
\newcommand{\zorigin}{\ensuremath{z_{\mathrm{origin}}}\xspace}
\newcommand{\abszorigin}{\ensuremath{|z_{\mathrm{origin}}|}\xspace}
\newcommand{\zpv}{\ensuremath{z_{\mathrm{PV}}}\xspace}
\newcommand{\tg}{\ensuremath{t_\gamma}\xspace}
\newcommand{\lowMET}{\ensuremath{\MET < 20~{\rm GeV}}\xspace}
\def\Wenu{\ensuremath{W \rightarrow e\nu}}

\usepackage{preprintcover}

\PreprintCoverPaperTitle{Search for non-pointing and delayed photons in the  diphoton and missing transverse momentum
final state in 8~TeV $pp$ collisions at the LHC using the ATLAS detector}

\PreprintIdNumber{CERN-PH-EP-2014-215}

\PreprintCoverAbstract{A search has been performed,
using the full data sample of 20.3~\ifb \ of 8~TeV proton--proton collisions collected in 2012 with the ATLAS detector
at the LHC, for photons originating from a displaced vertex due to the decay
 of a neutral long-lived particle into a photon and an invisible particle.
The analysis investigates the diphoton plus missing transverse momentum final state, and is therefore
most sensitive to pair-production of long-lived particles. The analysis technique
exploits the capabilities of the ATLAS electromagnetic calorimeter to make precise measurements
of the flight direction, as well as the time of flight, of photons.
No excess is observed over the Standard Model predictions for background.
Exclusion limits are set within  the context of Gauge Mediated Supersymmetry Breaking models,
with the lightest neutralino being the  next-to-lightest supersymmetric particle and decaying into a photon and gravitino
with a lifetime in the range from  250~ps up to about 100~ns.
}

\PreprintJournalName{Physical Review D}

\begin{document}

\title{Search for non-pointing and delayed photons in the  diphoton and missing transverse momentum
final state in 8~TeV $pp$ collisions at the LHC using the ATLAS detector}

\author{The ATLAS Collaboration}

\begin{abstract}
A search has been performed,
using the full 20.3~\ifb \ data sample of 8~TeV proton--proton collisions collected in 2012 with the ATLAS detector
at the LHC, for photons originating from a displaced vertex due to the decay
 of a neutral long-lived particle into a photon and an invisible particle.
The analysis investigates the diphoton plus missing transverse momentum final state, and is therefore
most sensitive to pair-production of long-lived particles. The analysis technique
exploits the capabilities of the ATLAS electromagnetic calorimeter to make precise measurements
of the flight direction, as well as the time of flight, of photons.
No excess is observed over the Standard Model predictions for background.
Exclusion limits are set within  the context of Gauge Mediated Supersymmetry Breaking models,
with the lightest neutralino being the  next-to-lightest supersymmetric particle and decaying into a photon and gravitino
with a lifetime in the range from  250~ps to about 100~ns.
\end{abstract}


\maketitle

\section{Introduction}

This article reports the results of a
search for photons originating from a displaced vertex due to the decay
of a neutral long-lived particle into a photon and an invisible particle.
The search exploits
the capabilities of the ATLAS liquid-argon (LAr) electromagnetic
(EM) calorimeter to make precise measurements
of the flight direction and  the time of flight of photons.
The analysis uses the full data sample of 8~TeV proton--proton ($pp$) collisions collected in 2012
with the ATLAS detector at the CERN Large Hadron Collider (LHC),
corresponding to an integrated luminosity of 20.3~\ifb. The method used is an evolution
of the ATLAS non-pointing photon analysis~\cite{Aad:2013oua} using the full 2011 data
sample of 7~TeV $pp$ collisions,
corresponding to an integrated luminosity of 4.8~\ifb. This previous analysis
based on 7~TeV $pp$ collisions found no excess above the Standard Model (SM) background expectation.

Scenarios where neutral long-lived particles are produced in pairs arise naturally, for example, within models of supersymmetry
(SUSY)~\cite{Miyazawa:1966, Ramond:1971gb, Golfand:1971iw, Neveu:1971rx, Neveu:1971iv, Gervais:1971ji, Volkov:1973ix, Wess:1973kz, Wess:1974tw}.
SUSY predicts the existence of a new SUSY partner (sparticle) for each of the SM particles,
with identical quantum numbers except differing by  half a unit of spin. In R-parity-conserving SUSY
models~\cite{Fayet:1976et,Fayet:1977yc,Farrar:1978xj,Fayet:1979sa,Dimopoulos:1981zb},
$pp$ collisions at the LHC could produce these sparticles in pairs, and they would then decay in cascades
involving other sparticles and SM particles until the lightest SUSY particle~(LSP) is produced, which is stable.
This analysis investigates the diphoton plus large \MET \ final state, where \MET \ is
the magnitude of the missing
transverse momentum, and is therefore most sensitive to the pair-production of long-lived particles.

In gauge-mediated supersymmetry breaking~(GMSB)
models~\cite{Dine:1981gu,AlvarezGaume:1981wy,Nappi:1982,Dine:1993yw, Dine:1994vc,Dine:1995ag},
the gravitino (\gravitino) is the LSP and is predicted, for typical model parameter values, to be very light.
While the recent discovery of a Higgs boson with a mass around 125~GeV~\cite{2012gk,Chatrchyan:2012ufa} disfavors minimal
GMSB within reach of the LHC, modifications to
minimal GMSB can easily accommodate this Higgs mass value without changing the sparticle masses~\cite{Abdullah:2012tq,Evans:2011bea,Evans:2012hg}.
GMSB phenomenology is largely determined by the properties of the
next-to-lightest supersymmetric particle~(NLSP), since the decay chains of the sparticles with higher
mass would terminate in the decay of the NLSP.
Very weak coupling of the NLSP to the gravitino could lead to displaced decay vertices of the NLSP~\cite{Dine:1994vc}.
The NLSP lifetime ($\tau$) depends on the
fundamental scale of SUSY breaking~\cite{Giudice:1998bp,Chen:1997yfa},
and therefore provides important information about the SUSY-breaking mechanism.

The results of this analysis are presented within the context of the so-called Snowmass Points and Slopes parameter set 8
(SPS8)~\cite{Allanach:2002nj}, which describes a set of minimal GMSB models with the lightest
neutralino (\ninoone) as the NLSP.  The free parameter in the GMSB SPS8 set of models
is the effective scale  of SUSY breaking, denoted $\Lambda$, which depends on details of
how the SUSY breaking is communicated to the messenger sector of the theory.

For $\Lambda$ values below about 100~TeV, strong production of
pairs of squarks and/or gluinos make a significant contribution to the production rate of SUSY events at the LHC.
However, for most of the range of $\Lambda$ values relevant for this analysis,
SUSY production is dominated by electroweak pair production
of gauginos, and in particular of \ninotwo \chinoonepm
\ and \chinoonep \chinoonem \ pairs.

In the GMSB SPS8 models,
the dominant decay mode of the NLSP is $\ninoone \ra \gamma + \gravitino$,
leading to a $\gamma\gamma + \MET + X$ final state, where the escaping gravitinos
give rise to \MET,   and $X$
represents SM particles produced in the decay cascades.
To minimize the dependence of the results on the
details of the SUSY decays, the analysis requires only a pair of photons and large \MET,
avoiding explicit requirements on the presence of leptons or jets or
any other particular SM particles in the final state.

This analysis considers the scenario where the NLSP has a finite lifetime, at least 250~ps, and
travels partway through the ATLAS detector before decaying. In the range of $\Lambda$ values of
interest, about 80--300~TeV, the NLSP mass lies in the range of about 120--440~GeV. In
this case, the photons produced in the NLSP decays can either be  ``non-pointing'' or ``delayed''
or both; namely, the photons can have flight paths that do not point back to the primary vertex (PV) of
the event and arrival times at the calorimeter that are later than those expected
for a photon produced promptly at the PV.

The search for non-pointing and delayed photons is performed using the excellent performance
of the finely segmented LAr EM calorimeters. An EM shower produced
by a photon is measured precisely with varying lateral segmentation in three different
longitudinal (i.e. depth) segments, allowing a determination of the flight direction of the photon
from the EM shower measurements. The flight direction can then be compared with
the direction back toward the PV identified for the event. This method is employed to determine the value of the
pointing-related variable used, namely \abszdca, defined as the separation, measured along the beamline,
between the extrapolated origin of the photon and the position of the selected PV of the event.
The LAr calorimeter also has excellent
time resolution and the arrival time \tg \ of a photon at the calorimeter (with zero defined
as the expected value for a prompt photon from the hard collision) is also a sensitive measure,
since positive and finite time values would be expected for
photons arising from non-prompt NLSP decays.

In the 7~TeV analysis~\cite{Aad:2013oua}, the pointing measurement was used to extract the result,
with the time measurement used only qualitatively as a cross-check. The 7~TeV analysis
set exclusion limits within the context of GMSB SPS8 models and similar results were obtained in
a CMS analysis~\cite{Chatrchyan:2012jwg} of their full 7~TeV dataset, but investigating a final
state with at least one photon, at least three jets, and \MET.
The current analysis utilizes both the pointing and time measurements. As described in
Sec.~\ref{sec:stats}, the current analysis divides the sample into six exclusive categories,
according to the value of \abszdca, and then simultaneously fits the \tg \ distributions of
each of the categories to determine the possible contribution from signal. The use
of both variables greatly improves the sensitivity.

\section{The ATLAS Detector}
\label{sec:detector}

 The ATLAS detector~\cite{Aad:2008zzm} covers nearly the entire solid
 angle\footnote{ATLAS uses a
right-handed coordinate system with its origin at the nominal interaction point (IP)
in the center of the detector and the $z$-axis along the beam pipe. The $x$-axis
points from the IP to the center of the LHC ring, and the $y$-axis points upward.
Cylindrical coordinates $(r,\phi)$ are used in the transverse plane, $\phi$ being
the azimuthal angle around the beam pipe. The pseudorapidity is defined in terms
of the polar angle $\theta$ as $\eta=-\ln\tan(\theta/2)$, and
the transverse energy as $\ET = E \sin\theta$.}
around the collision point and consists of an inner tracking detector surrounded
 by a solenoid, EM and hadronic calorimeters,
 and a muon spectrometer incorporating three large toroidal magnet systems.
The inner-detector system (ID) is immersed in a 2~T axial magnetic field, provided
by a thin superconducting solenoid located before the calorimeters, and
provides charged-particle tracking in the pseudorapidity range $|\eta| < 2.5$.
The ID consists of three detector subsystems, beginning closest to the beamline
with a high-granularity silicon pixel detector,
followed at larger radii by a silicon microstrip tracker and then a straw-tube-based
transition radiation tracker.  The ID allows an accurate reconstruction of tracks from the primary $pp$
collision and precise determination of the location of the PV.

This analysis relies heavily on the capabilities of the ATLAS calorimeter system, which
covers the pseudorapidity range $|\eta|< 4.9$. Finely segmented lead/LAr EM sampling
calorimeters cover the  barrel ($|\eta| < 1.475$) and endcap ($1.375 < |\eta| < 3.2$) regions. An additional thin LAr
presampler covering $|\eta| < 1.8$ allows corrections for energy losses in material upstream of the EM calorimeters.
Hadronic calorimetry is provided by a steel/scintillator-tile calorimeter, segmented into three barrel
structures within $|\eta| < 1.7$, and two copper/LAr hadronic endcap calorimeters. The solid
angle coverage is completed with forward copper/LAr and tungsten/LAr calorimeter modules,
optimized for EM and hadronic measurements, respectively.  Outside the calorimeters lies the
muon spectrometer, which identifies muons and measures their deflection up to $|\eta| = 2.7$
 in a magnetic field generated by superconducting air-core toroidal magnet systems.

 \subsection{Pointing resolution}

 For $\abseta < 2.5$,
 the EM calorimeter is segmented into three layers in depth that are used to measure the longitudinal profile
of the shower. The first layer uses highly
 granular ``strips'' segmented in the $\eta$ direction, designed to allow efficient discrimination between single
 photon showers and two overlapping showers, the latter originating, for example, from the decay of a $\pi^0$ meson. The second layer
collects most of the energy deposited in the calorimeter by EM showers initiated by electrons or photons.
Very high energy showers can leave significant energy deposits in the third layer, which can also
 be used to correct for energy leakage beyond the EM calorimeter.

By measuring precisely the centroids of the EM shower in the first and second EM calorimeter layers,
the flight direction of photons can be determined, from which one can calculate the value of
\zorigin, defined as the $z$-coordinate of the photon projected back to the point giving its
distance of closest approach
to the beamline ($x = y = 0$).
The angular resolution of the EM calorimeter's measurement of the flight direction of prompt photons
is about 60~mrad/$\sqrt{(E/ {\rm GeV})}$, where $E$ is the photon energy.
This angular precision
corresponds, in the EM barrel calorimeter, to a resolution in \zorigin \ of about 15~mm for prompt photons with
 energies in the range of 50--100~GeV. Given the geometry, the
$z$ resolution is worse for photons reconstructed in the endcap calorimeters, so the pointing analysis is
restricted to photon candidates in the EM barrel calorimeter.

In the ATLAS $H \ra \gamma\gamma$ analysis~\cite{2012gk} that contributed to the discovery of a
Higgs boson, this capability of the EM calorimeter was used to help choose the PV
from which the two photons originated, thereby improving the diphoton invariant mass resolution and
the sensitivity of the search.  The analysis described in this paper uses the measurement of the
photon flight direction to search for photons that do not point back to the PV. The pointing variable used in the
analysis is \zDCA, defined as the difference between \zorigin \ and \zpv, the $z$-coordinate
of the selected PV of the event.
Given that \zpv \ is measured with high precision using the tracker, the
\zorigin \ resolution is essentially equivalent to the resolution in \zdca.

While the geometry of the EM calorimeter is optimized for detecting particles that point back to near
the nominal interaction point at the center of the detector (i.e. $x = y = z = 0$), the fine segmentation allows good
pointing performance to be achieved over a  wide range of photon impact angles.
Figure~\ref{fig:zdcaresn}  shows the expected pointing resolution  (ie. the resolution of the measured \zorigin)
as a function of \abszorigin,
for GMSB SPS8 signal photons in the EM barrel calorimeter.
The results are obtained from Monte Carlo (MC) simulations (see Sec.~\ref{sec:mc:simulation}) by fitting to a Gaussian function
the difference between the values of \zorigin \
obtained from the calorimeter measurement and the MC generator-level information.
The pointing resolution degrades with increasing \abszorigin,
but remains much smaller than \abszorigin \ in the region where the signal is expected.

The calorimeter pointing performance was
verified in data by using the finite spread of the LHC collision region along the $z$-axis.
The pointing resolution achieved for a sample of electrons from \Zee \ events is also shown
in Fig.~\ref{fig:zdcaresn},
where the distance, \zpv, between the PV and the nominal center of the detector
serves the role of \zorigin. In this case, the pointing resolution is obtained
by fitting to a Gaussian the difference between \zpv, obtained from reconstructed tracks,
and the calorimeter measurement of the origin along the beamline of the electron.
Figure~\ref{fig:zdcaresn} shows that a similar
pointing performance is observed for photons and for electrons, as expected given their similar EM shower developments.
This similarity validates the use of a sample of electrons from \Zee \ events to study the
pointing performance for photons. The expected pointing performance for electrons in a MC sample of
\Zee \ events is also shown on Fig.~\ref{fig:zdcaresn}, and is consistent with the data.
  The level of agreement between MC simulation and data over the range of values that can be accessed
in the data gives confidence in the extrapolation using MC simulation to the larger \abszorigin \ values characteristic
of signal photons.

\begin{figure}[hbpt]
\centering
\includegraphics[width=0.48\textwidth]{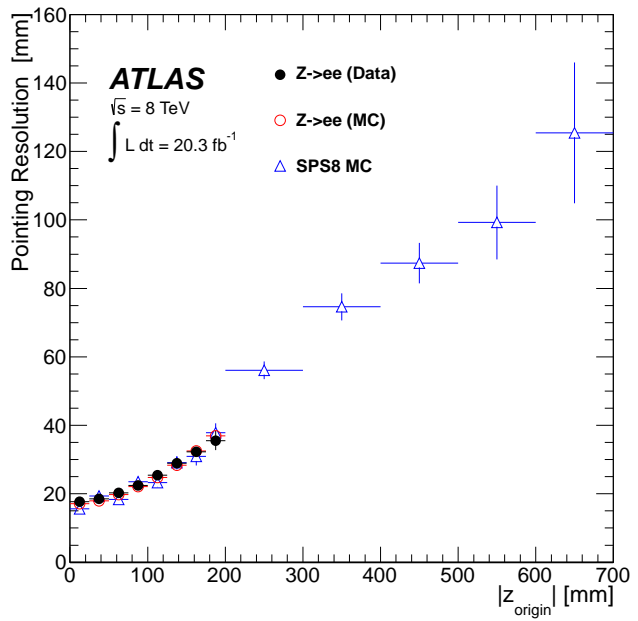}
\caption{The pointing resolution (defined as the resolution of \zorigin) obtained for EM showers in the LAr EM barrel calorimeter.
The pointing resolution for photons from GMSB SPS8 signal MC samples is plotted as a
function of \abszorigin. The pointing resolution is also shown for \Zee \ data and MC events,
for which the PV position, \zpv, serves the role of \abszorigin.}
\label{fig:zdcaresn}
\end{figure}

\subsection{Time resolution}
\label{sec:timing}

Photons from long-lived NLSP decays would reach the LAr calorimeter with
a slight delay compared to prompt photons produced directly in the hard scatter. This
delay results
mostly from the flight time of the heavy NLSP, which would have  a  distribution
of relativistic speed ($\beta = v/c$) that peaks typically near 0.9 and
has a tail to much lower values.  In addition,
the opening angle in the NLSP decay, which causes the photon to be non-pointing,
results in a longer geometrical path to the calorimeter, as compared to a prompt photon from the PV.

The EM calorimeter, with its novel ``accordion'' design, and its readout, which
incorporates fast shaping, has excellent time resolution.
Quality-control tests during production of
the electronics required the clock jitter on the LAr readout boards to be less than 20~ps,
with typical values of 10~ps~\cite{Buchanan:2008zza}. Calibration tests of the overall
electronic readout performed {\sl in situ} in the ATLAS cavern show a time resolution
of $\approx 70$~ps~\cite{Abreu:2010zzc}, limited not by the readout but by the jitter of
the calibration pulse injection system. Test-beam measurements~\cite{2008NIMPA.597..178A}
of EM barrel calorimeter modules demonstrated a time resolution
of $\approx 100$~ps in response to high-energy electrons.

The LAr energy and time for each calorimeter cell
are reconstructed by applying the
optimal filtering algorithm~\cite{cleland} to the set of five samples of the signal shape
read out for each calorimeter channel,
with successive samples on the waveform separated by 25~ns.  More specifically,
the deposited energy per cell and the time of the deposition are calculated using
appropriately weighted linear combinations of the set of samples of the waveform:

\begin{equation}
\label{eq:energy}
E  = \sum_{i=0}^{4} a_i S_i  \ \ \    {\rm and} \ \ \
t  = \frac{1}{E} \sum_{i=0}^{4} b_i S_i  \ \     ,
\end{equation}

\noindent where $S_i$
denotes the five samples of the signal waveform. The parameters
$a_i$ and $b_i$ are the optimal filter coefficients (OFC), the values of which are
calculated, knowing the pulse shape and noise autocorrelation matrix, to deliver the
best energy and time resolutions.

For this analysis, the arrival time of an
EM shower is measured using the second-layer EM calorimeter cell with the
maximum energy deposit.  For the EM shower of an electron or photon with energy
within the range of interest, this cell typically
contains about (20--50)\% of the total energy deposited in the EM shower.
In principle, the times measured in neighboring cells could be used in a weighted
time calculation to try to further improve the precision. However, some studies that investigated more
complicated algorithms found no improvement in time resolution, likely due to the pulse shapes in the
channels with lower deposited energies suffering some distortion due to crosstalk effects.

During 2012,
the various LAr channels were timed-in online with a precision of order 1~ns.
A large sample of $W \rightarrow e\nu$  events in the 8~TeV dataset was used
to determine calibration corrections that need to be applied to optimize the
time resolution for EM clusters. The calibration includes corrections of
various offsets in the time of individual channels, corrections for the energy
dependence of the time measurement, crosstalk corrections, and flight-path corrections depending on the PV
position.

To cover the full dynamic range of physics signals of interest, the
ATLAS LAr calorimeter readout boards~\cite{Buchanan:2008zza} employ three
overlapping linear gain scales, dubbed High, Medium and Low, where
the relative gain is reduced by a factor of about ten for each successive scale.  For a given event,
any individual LAr readout channel is digitized using the gain scale that provides
optimal energy resolution, given the energy deposited in that calorimeter cell.
The calibration of the time was determined separately for High and
Medium gain for each channel. The number of electron candidates from
the \Wenu \ sample that were digitized using Low gain was insufficient to obtain
statistically precise results for the calibration constants. Therefore, the analysis
requires that selected photons be digitized using either High or Medium gain
resulting in a loss in signal efficiency, which ranges from much less than 1\%,
for the lowest $\Lambda$ values probed, to less than 5\% for the highest $\Lambda$ values.
The majority of signal photons are digitized
using Medium gain, the fraction rising with rising $\Lambda$ from about 60\% to about 90\%,
 over the $\Lambda$ range of interest.

An independent  sample
of \Zee \ events was used to validate the time calibration and determine the
resolution obtained, by performing Gaussian fits to the time distributions in bins
of cell energy.
Figure~\ref{fig:timingresn} shows the time resolution for High and Medium
gain cells with $|\eta| < 0.4$, as a function of the energy in the second-layer
calorimeter cell used to calculate the time for the sample of
\Zee \ events.  Similar results are obtained
over the full coverage of the EM calorimeter.

\begin{figure}[hbpt]
\centering
\includegraphics[width=\columnwidth]{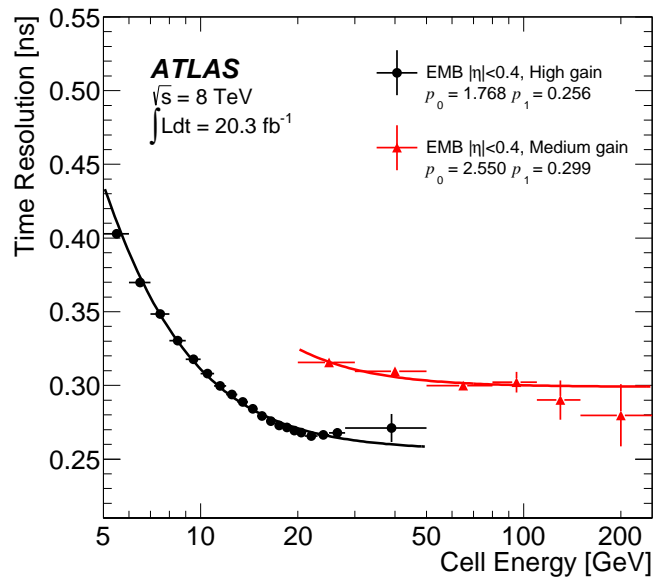}
\caption{Time resolution, as a function of the energy in the second-layer
cell with the maximum energy, obtained from \Zee \ events, for electrons
in the EM barrel calorimeter (EMB) with $|\eta| < 0.4$, and for both the
High and Medium gains. Similar results are obtained over the full
coverage of the EM calorimeter. The energy deposited in this cell
is typically about (20--50)\% of the total energy of the electron.
Included in the figure are the results of fitting the time resolution
results to the expected form of $\sigma(t) = p_0/E \oplus p_1$,
with fit parameters $p_0$ ($p_1$) measured in units
of GeV$\cdot$ns (ns). The time resolution includes a contribution of $\approx 220$~ps,
which is due to the LHC bunch-spread along the beamline.}
\label{fig:timingresn}
\end{figure}

The time resolution, $\sigma(t)$, is expected to follow
the form $\sigma(t) = p_0/E \oplus p_1$, where $E$ is the cell energy,
$\oplus$ indicates addition in quadrature, and the fit parameters
$p_0$ and $p_1$ are the coefficients of the so-called noise term
and constant term, respectively.
Superimposed on Fig.~\ref{fig:timingresn} are the results of fits to
this expected form of the time resolution function. The fits yield
values of $p_1$, which gives the time
resolution in the limit of large energy deposits, of 256~ps (299~ps)  for
High (Medium) gain.
The somewhat worse results for Medium gain are due to limited statistics in
the \Wenu \ sample used to determine the time calibration constants.
The time resolution includes a contribution of $\approx 220$~ps,
which is caused by the time spread in $pp$ collisions for a given PV position due to the LHC
bunch-spread
along the beamline. Subtracting this contribution in quadrature
implies the LAr contributions to the time resolution are
$\approx 130$~ps ($\approx 200$~ps) for High (Medium) gain.

The time resolution is not modeled properly in the MC simulation of the ATLAS detector and
it is necessary to apply additional smearing to the MC events in order to match
the time performance observed in data.   To smear the MC events, the fits to the time resolution
determined from \Zee \ data as a function of the energy of the most energetic cell in the second layer are
used. The fits are parameterized in terms of the pseudorapidity of the cell and the
gain scale used to reconstruct the time. To account for the impact of the
beam-spread, the smearing includes a component with a Gaussian
standard deviation of 220~ps that is applied in a correlated way to all photons
in the same event. In addition, an uncorrelated component is
applied separately to each photon to match its overall time resolution to that observed in data.

\subsection{Measurements of delayed particles}
\label{sec:delayed}

The OFC values in Eq.~\ref{eq:energy}
deviate from being optimal for signals that are
early or delayed with respect to the time used to determine the OFC values.
This effect can cause the reconstructed values of the energy and time to deviate from their
true values.

A source of early and delayed particles can be obtained using so-called satellite bunches of protons
that, due to the radio-frequency structure of the LHC accelerator and injection complex, are present
in the LHC beams but separated from the main bunches by multiples of $\pm 5$~ns.
A study was made using \Wenu \ and \Zee \ events produced
in collisions between pairs of such satellite bunches that occur at the center of the
detector but are 5 ns early or late, compared to nominal collisions. These ``satellite--satellite'' collisions
are suppressed in rate by a factor of about one million compared to collisions of the
nominal bunches, since the typical population of a satellite bunch is about a factor of
one thousand lower than that of the nearby nominal bunch.  However, the 8~TeV data sample is
sufficiently large that a statistically significant observation of these satellite--satellite
collisions could be made.

The values of the mean times reconstructed for electrons produced in
satellite--satellite collisions were determined to be $\approx-5.1$~ns ($\approx+5.4$~ns),
for events that occurred 5~ns early (late), demonstrating that the use of fixed
OFC values causes a bias for signals that are sufficiently early or late compared
to the nominal time. In contrast to the time reconstruction, the studies show that the reconstructed energies are
very insensitive to modest time shifts of the samples on the waveform,
as expected due to the methods used to calculate
the OFC values used in the energy calculation. For time shifts within $\pm 5$~ns
of the nominal time, the reconstructed energy decreases by less than 1\%.

\section{Data and Monte Carlo Simulation Samples}
\label{sec:mc:simulation}

This analysis uses  the full dataset of $pp$ collision events at a center-of-mass
energy of $\sqrt{s}=\unit[8]{TeV}$, recorded  with the ATLAS detector in 2012. The
data sample, after applying quality criteria that require all ATLAS sub-detector
systems to be functioning normally, corresponds to a total integrated luminosity of $20.3~\ifb$.

While all background studies, apart from some
cross-checks, are performed with data, MC simulations are
used to study the response to GMSB signal models, as a function
of the free parameters $\Lambda$ and $\tau$.  The other GMSB parameters are fixed to the following
SPS8 model values: the messenger mass $M_{\rm mess} =2\Lambda$, the number
of SU(5) messengers $N_5 = 1$, the ratio of the vacuum expectation
values of the two Higgs doublets $\tan \beta = 15$,
and the Higgs-sector mixing parameter $\mu > 0$~\cite{Allanach:2002nj}.

The full GMSB SPS8 SUSY mass spectra, branching fractions and decay widths
are calculated from this set of parameters using {\tt ISAJET}~\cite{isajet} version~7.80.
The {\tt HERWIG++} generator, version 2.4.2~\cite{Bahr:2008pv}, was used to generate the signal MC samples,
 with MRST\,2007\,LO${}^{*}$~\cite{Sherstnev:2007nd} parton density distributions (PDF). A total of 30 signal points, from $\Lambda=\unit[70]{TeV}$ to
$\Lambda=\unit[400]{TeV}$, were generated, with  $\tau$ values of 2~ns  or 6~ns. For each signal point,  40,000 inclusive GMSB SUSY events
were simulated. For each sample, the NLSP was forced to decay to a photon and gravitino, with the branching
fraction $BR(\ninoone \ra \gamma \gravitino)$ fixed to unity. Other $\tau$ values were simulated by appropriately
reweighting the events of these generated samples, with weights related to the decay times of the neutralinos, to
mimic the expected decay time distributions.

 Signal cross sections are calculated to next-to-leading order (NLO) in the strong coupling constant
  using PROSPINO2~\cite{Beenakker:1996ch}\footnote{In addition a  resummation
 of soft gluon emission at next-to-leading-logarithm accuracy
 (NLL)~\cite{Beenakker:1996ch,Kulesza:2008jb,Kulesza:2009kq,Beenakker:2009ha,Beenakker:2011fu}
 is performed in the case of strong SUSY pair production.}. The nominal cross section
 and its uncertainty are taken from an envelope of cross-section predictions using different
 PDF sets and factorization and renormalization scales, as described in \Ref~\cite{Kramer:2012bx}.
 Uncertainties on the cross-section values range from 9\% to 14\%.

All MC samples used in this analysis were passed through a {\tt
  GEANT4}-based simulation~\cite{geant4,Aad:2010wqa} of the ATLAS
detector and were reconstructed with the same
algorithms used for the data.  The effect of multiple
$pp$ interactions in the same or nearby bunch
crossings (pileup) is taken into account in all MC simulations
and the distribution of the number of interactions
per bunch crossing in the MC simulation is reweighted
to that observed in the data. During the  2012 data-taking period, the
average number of $pp$ collisions per bunch crossing varied between 6 and 40, with a mean value of 20.7,

\section{Object Reconstruction and Identification}

The reconstruction and identification of electrons and photons are described in
Refs.~\cite{Aad:2011mk, Aad:2014fxa} and ~\cite{Aad:2010sp}, respectively.
The photon identification criteria described in Ref.~\cite{Aad:2010sp}
have been re-optimized for the expected pileup conditions of the 8~TeV run period.
Shape variables computed from the lateral and
longitudinal energy profiles of the EM showers in the calorimeter are used to identify
photons and discriminate against backgrounds.  A set of photon selection criteria, designed
for high efficiency and modest background rejection, defines the so-called ``{\it loose}''
photon identification used in this analysis. The loose photon requirements
use variables that describe the shower shape in the second layer of the EM
calorimeter and leakage into the hadronic calorimeter.  These selection criteria do not
depend on the transverse energy of the
photon (\ET), but do vary as a function of $\eta$ in order to take into account variations
in the calorimeter geometry and upstream material.  The efficiency of these loose requirements,
for the signal photons, is over 95\% over the range $\abszorigin<250$~mm and steadily
falls to approximately 75\% at $\abszorigin=700$~mm.

The measurement of \met \ ~\cite{Aad:2012re} is based on the energy deposits in the calorimeter
with $|\eta|<4.9$ and the energy associated with reconstructed muons; the latter is estimated
using the momentum measurement of its reconstructed track.
The energy deposits associated with reconstructed objects (jets defined using the
anti-$k_t$ algorithm~\cite{Cacciari:2008gp} with radius parameter 0.4, photons,
electrons) are calibrated accordingly. Energy deposits not associated with a reconstructed object are
calibrated according to their energy sharing between the EM and hadronic calorimeters.

\section{Event Selection}

The selected events were collected by an online trigger requiring the presence of at least two loose photons
with $|\eta|< 2.5$, one with $\ET> 35$ GeV and the other with $\ET> 25$ GeV.
This trigger is insensitive to the time of arrival of photons that are relevant for the signal considered, but there
may be a slight dependence of the trigger efficiency on the \zorigin of the photon.  This effect is discussed
in Sec.~\ref{sec:sigyield}.  The trigger efficiency exceeds 99\% for signal events that pass the offline selection cuts.
To ensure the selected events resulted from a $pp$ collision, events
are required to have at least one PV candidate with five
or more associated tracks, each with transverse momentum satisfying $\pT > 400$~MeV. In case of multiple vertices,
the PV is chosen as the vertex with the greatest sum of
the squares of the transverse momenta of all associated tracks.

The offline photon selection requires two loose photons with $\ET> 50$ GeV and  $|\eta|< 2.37$
(excluding the transition region between the barrel and endcap EM calorimeter at $1.37<|\eta|< 1.52$).
At least one photon is required to be in the barrel region  $|\eta|< 1.37$.  Both photons are
required to be isolated, by requiring that the transverse energy deposited in the calorimeter in a cone of radius
$\Delta R=\sqrt{(\Delta\eta)^2+(\Delta\phi)^2}=0.4$ around each photon candidate be less than 4 GeV,
after corrections to account for pileup and the energy deposition from the
photon itself~\cite{Aad:2010sp}. To avoid collisions due to satellite bunches, both photons are required to have
a time that satisfies $|\tg| < 4$~ns.

The selected diphoton sample is divided into exclusive subsamples according to the value of \met.
The subsample with \met$<20$ GeV is used to model the prompt backgrounds, as described in
Sec.~\ref{sec:bgtemplates}. The events with 20 GeV $<\met<$ 75 GeV are used as control
samples to validate the analysis procedure and background model.
Diphoton events with  $\met>$ 75 GeV define the signal region.

Table~\ref{tab:acc} summarizes the total acceptance times efficiency of the selection requirements
for examples of GMSB SPS8 signal model points with various $\Lambda$ and $\tau$ values.  Strong SUSY
pair production is only significant for $\Lambda<$100 TeV.  For $\Lambda=80$ TeV and $\tau=6$ ns,
the acceptance times efficiency is evaluated from MC samples to be $1.6\pm0.1$\% and $2.1\pm0.1$\% for
weak and strong production, respectively, corresponding to a total value of $1.7\pm0.1$\%.
For fixed $\Lambda$, the acceptance falls approximately exponentially with increasing $\tau$, dominated by the requirement
that both NLSP decay before reaching the EM calorimeter,
so that the resulting photons are detected.
For fixed $\tau$, the acceptance increases with increasing $\Lambda$,
since the SUSY particle masses increase, leading the decay cascades to produce, on average, higher  \MET \
and also higher \ET \ values of the decay photons.

\begin{table}[htbp]
\begin{center}
\caption{The total signal acceptance times efficiency, given in percent, of the event selection requirements,
 for sample GMSB SPS8 model points with various $\Lambda$ and $\tau$ values. The uncertainties
 shown are statistical only.\label{tab:acc} }
\begin{tabular}{cccc}          \hline	\hline
$\tau$   &  \multicolumn{3}{c}{Signal acceptance times efficiency [\%]} \\ \cline{2-4}
 [ns] & $\Lambda = 80$~TeV & $\Lambda = 160$~TeV  & $\Lambda = 320$~TeV  \\ \hline
 \ \   0.5 \ \   & $8.4 \pm 0.6$ \ &  $30 \pm 1$ \ & $46 \pm 2$   \   \\
  2 & $5.1 \pm 0.3$ \ & $21 \pm 0.2$ \ & $33.0 \pm 0.3$ \  \\
  6 & $1.7 \pm 0.1$ \ & $7.3 \pm 0.1$ \ & $12.5 \pm 0.2$ \  \\
  10 & $0.86 \pm 0.03$ & $3.71 \pm 0.06$ & $6.45 \pm 0.09$ \\
  40 & $0.089 \pm 0.004$ & $0.38 \pm 0.01$ & $0.70 \pm 0.02 $\\
  100 & \ $0.016 \pm 0.001$ \ & \ $0.070 \pm 0.002$ \ & \ $0.129 \pm0.004$ \ \\ \hline \hline
\end{tabular}
  \end{center}
\end{table}

\section{Signal and Background Modeling}
\label{sec:templates}

The analysis exploits both the pointing and time measurements.  However,
the measured properties of only one of the two photons are
used, where the choice of which photon to use is made according
to the location of the two photons. The selection requires at least one of the
photons to be in the barrel region, since events with both photons in the endcap calorimeters are expected to
contribute very little to the signal sensitivity.
For events, referred to hereafter as BE events, where
one photon is found in the barrel and
one in the endcap calorimeter, the \zdca \  and \tg \  measurements of the barrel photon are
used in the analysis; this choice is made since, due to geometry, the \zdca \  resolution in the
barrel calorimeter is better. For so-called BB events, with
both photons in the EM barrel calorimeter, the \zdca \ and \tg \
measurements of the photon with the maximum value of \tg \ are used. Studies showed
that this approach achieves a sensitivity very similar to that when using both photons, while
avoiding the complexity of having to deal with the correlations between the measurements
of the two photons within a single event.

\subsection{GMSB SPS8 signal}

The shape of the \zdca \ and \tg \ distributions for signal events is obtained from the
signal MC samples. For a given value of $\Lambda$,
the distributions for any NLSP lifetime
value can be obtained by appropriately reweighting the distributions of the existing MC samples.

Examples of  \zdca \ and \tg \ signal distributions for a few representative GMSB SPS8 models are
shown in Fig.~\ref{fig:signaltiming}.  The distributions are normalized to unity area within the
displayed horizontal-axis range, in order to allow for an easier comparison between the various
signal and background shapes. The upper two plots show signal shapes for  some
example NLSP lifetime  ($\tau$) values, all
with $\Lambda$ fixed to a value of $160$~TeV. The lower two plots show signal shapes for
some example $\Lambda$ values, all with $\tau$ fixed to a value of $1$~ns.
The signal shapes have some dependence on $\Lambda$ due
to its impact on the SUSY mass spectrum, and therefore the event kinematics. However,
the signal shapes vary most strongly with NLSP lifetime. For larger $\tau$ values,
the signal shapes are significantly impacted by the diphoton event selection, which
effectively requires that both NLSP decay before reaching the EM calorimeters, leading to
a signal acceptance that falls rapidly with increasing time values.  As a result, the
signal shapes for $\tau$ values of $2.5$~ns and $25$~ns, for example, are quite similar, as shown
in the upper plots of Fig.~\ref{fig:signaltiming}.

\begin{figure*}[hbpt]
\centering
\includegraphics[width=0.48\textwidth]{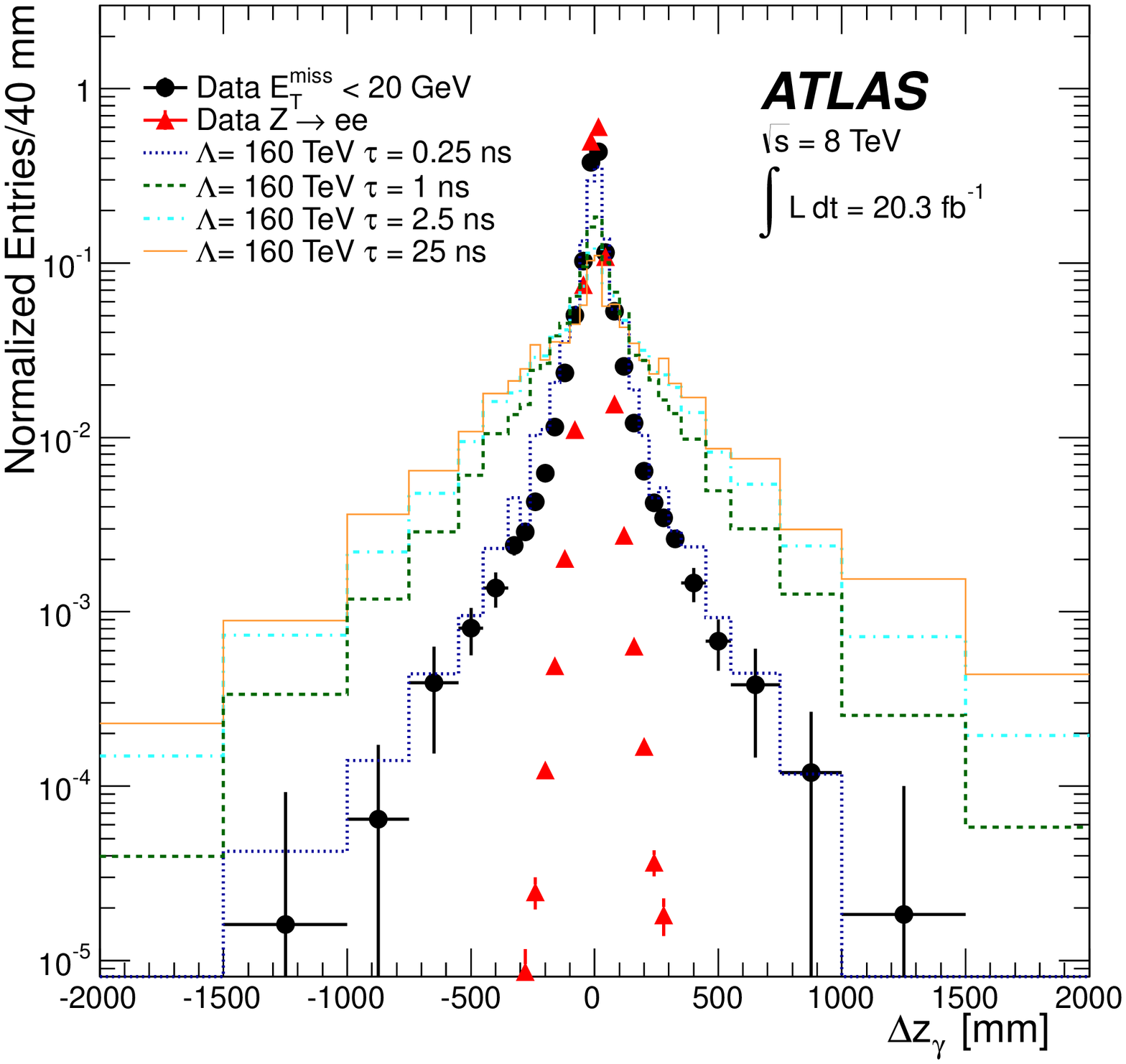}
\includegraphics[width=0.48\textwidth]{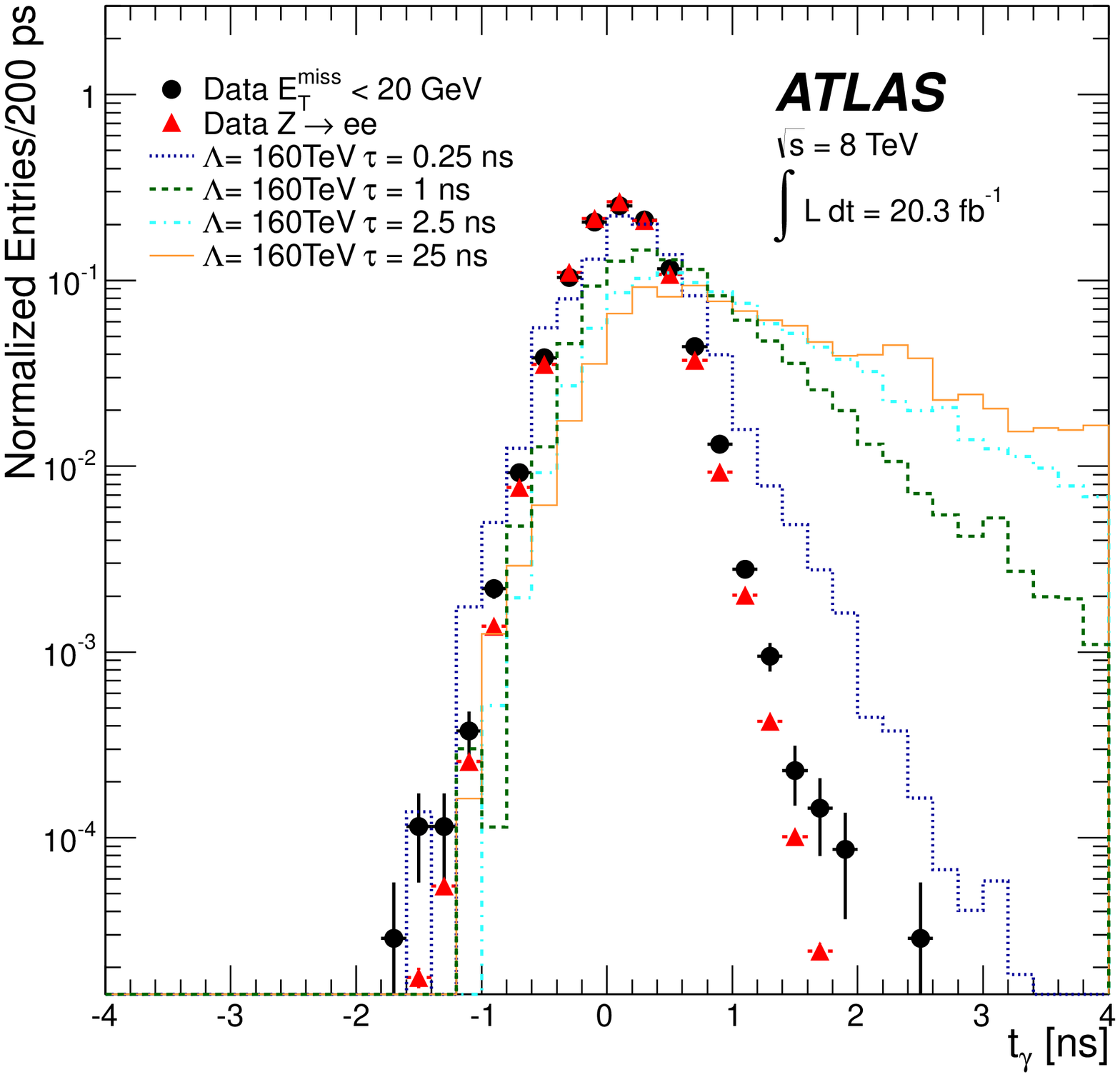}
\includegraphics[width=0.48\textwidth]{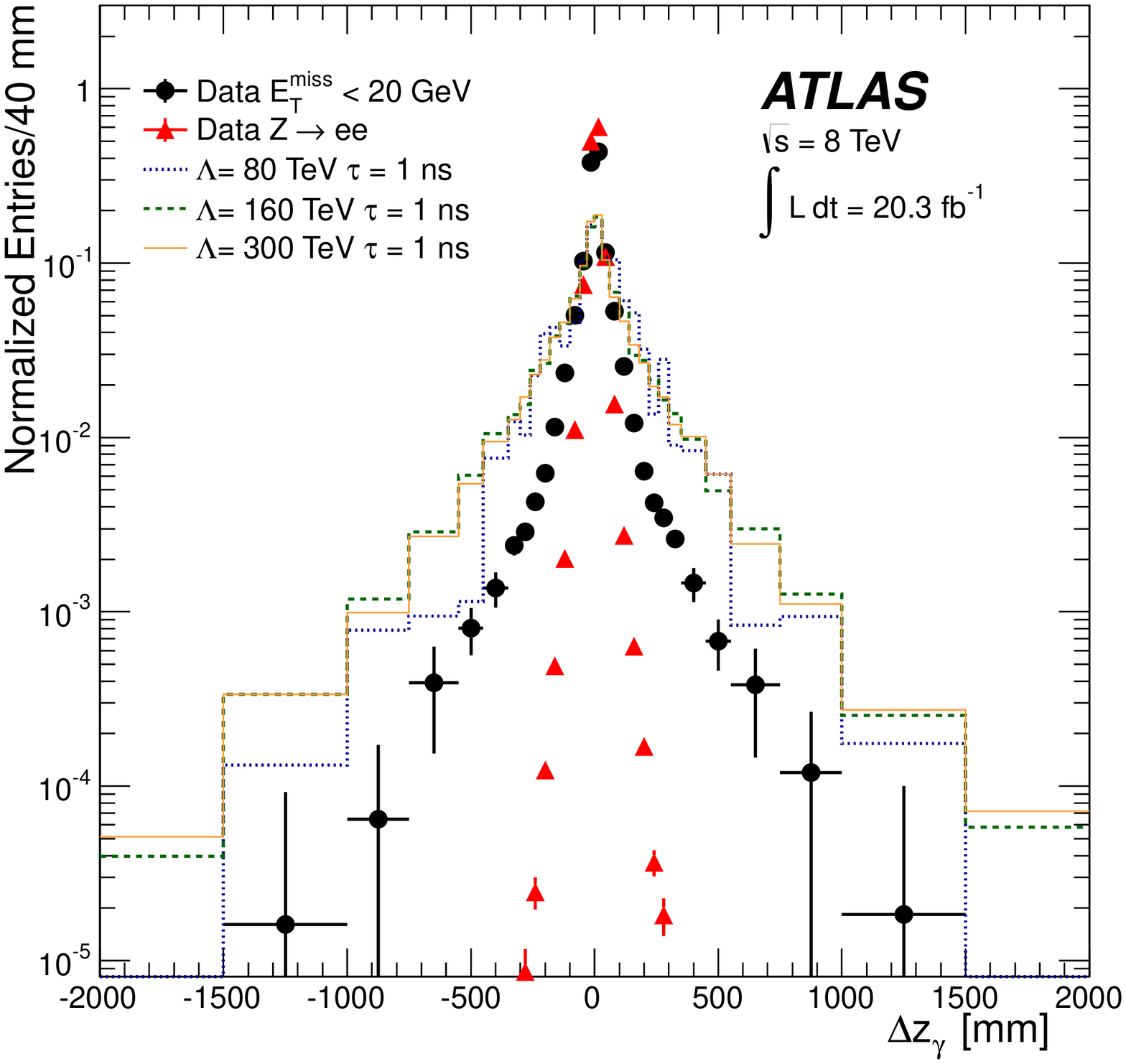}
\includegraphics[width=0.48\textwidth]{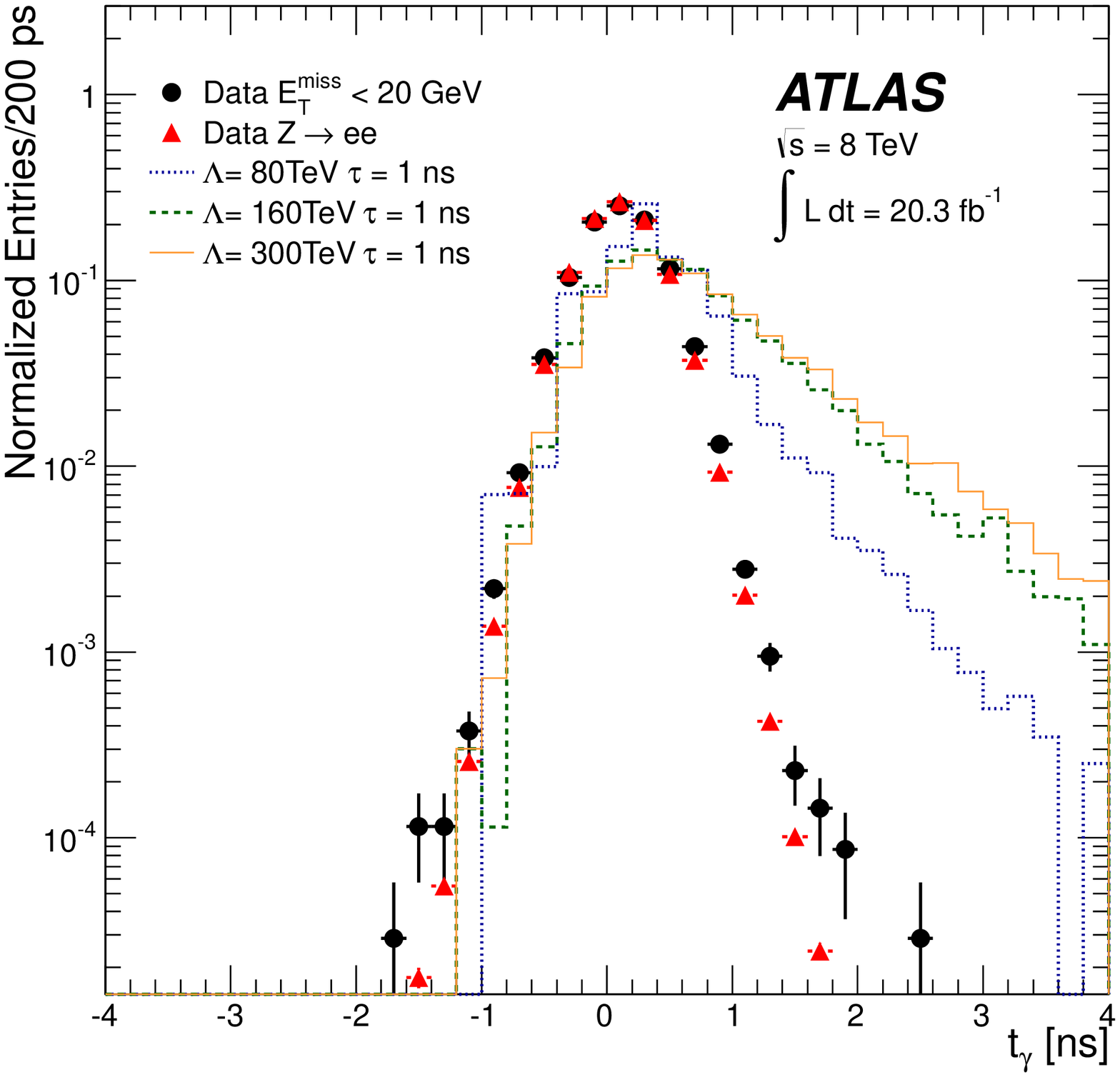}
\caption{Signal distributions for (left) \zdca \ and (right) \tg, for some example GMSB SPS8 model points.
The upper two plots show signal shapes for  NLSP lifetime  ($\tau$) values of $0.25, 1, 2.5, {\rm \ and \ } 25$~ns, all
with the effective scale of SUSY breaking
($\Lambda$) fixed to a value of $160$~TeV. The lower two plots show signal shapes for  $\Lambda$ values
of $80, 160, {\rm \ and \ } 300$~TeV, all with $\tau$ fixed to a value of $1$~ns.
Superimposed on each of the plots
are the corresponding data distributions for the samples used to
model the backgrounds, namely  \Zee \ events and diphoton events with $\MET < 20$~GeV.
For all plots, the distributions are normalized to unity area within the horizontal-axis range displayed,
and the uncertainties shown on the data distributions are statistical only.}
\label{fig:signaltiming}
\end{figure*}

\subsection{Backgrounds}
\label{sec:bgtemplates}

The background is expected to be completely dominated
by $pp$ collision events, with possible backgrounds due to cosmic rays, beam-halo events,
or other non-collision processes being negligible.
The source of the loose photons in background events contributing to the selected sample
is expected to be either a prompt photon, an electron misidentified as a photon, or a jet misidentified as a photon. In
each case, the object providing the loose photon signature originates from the PV.

The pointing and time distributions expected for the background sources are determined using
control samples in data. In addition to avoiding a reliance on the precise MC simulation of the pointing
and timing performance for the backgrounds, and particularly of the tails of their \zdca  and \tg distributions,
using data samples naturally accounts for the influence of pile-up, the possibility of selecting the wrong PV, and
any instrumental or other effects that might influence the background measurements.

Given their similar EM shower developments, the pointing and time resolutions for prompt photons are similar to those for electrons.
The \tg \ distribution in each \zdca category is modeled using electrons from $Z\ra ee$ data events.
The $Z\ra ee$ event selection requires a pair of oppositely charged electron candidates, each of which
has $\pT > 35$~GeV and $|\eta| < 2.37$ (excluding the transition region between the barrel
and endcap calorimeters).  Both electrons are
required to be isolated, with the transverse energy deposited in the calorimeter in a cone of size
$\Delta R=0.2$ around each electron candidate being less than 5 GeV, after subtracting the energy
associated with the electron itself. As for photons, electrons
must be read out using either High or Medium gain, and must have a time less than 4~ns.
The dielectron invariant mass is required to be within 10~GeV of the $Z$ boson mass,
yielding a sufficiently clean sample of $Z \ra ee$ events.
The electrons are used to construct \zdca and \tg templates.
The unit-normalized \Zee \ templates are shown superimposed on the plots of Fig.~\ref{fig:signaltiming}.

Due to their wider showers in the EM calorimeter,  jets have a
wider \zdca \ distribution than prompt photons and electrons. Events passing the diphoton selection with $\MET < 20~$GeV are used as a
data control sample that
includes jets with properties similar to the background contributions expected
in the signal region.  The \MET \ requirement  serves to render negligible any possible signal
contribution in this control sample. The time resolution depends on the deposited energy in the calorimeter.
Using the shape of the \lowMET \ template to describe
events in the signal region, defined with $\met > 75$~GeV therefore implicitly relies on the
kinematic distributions for photons in both regions being similar.  However, it is expected that there
should be a correlation between the value of \MET \ in a given event, and the \ET \
distribution of the physics objects in that event.  This correlation is indeed observed in the low-\MET\
control region samples. Increasing to 60~GeV the minimum \ET \ requirement on the photons in
the \lowMET \ control sample selects photons with
similar kinematic properties to the photons in the signal region. Therefore, the \lowMET
sample requiring $\ET > 60$~GeV for the photons is used to model the background.

The selected diphoton sample with $\MET < 20$~GeV should be
dominated by  jet--jet, jet--$\gamma$ and $\gamma\gamma$ events. Therefore,
the associated \zdca \ and \tg \ distributions include contributions from photons as well
as from misidentified jets that satisfy the loose photon signature.
The unit-normalized \lowMET \ templates are shown superimposed on the
plots of Fig.~\ref{fig:signaltiming}. As expected, Fig.~\ref{fig:signaltiming} shows that the \zdca \ distribution is much wider
for the \lowMET \ sample than for the \Zee \ sample, while the \tg \ distributions of
these two background samples are very similar.  Both
backgrounds have distributions that are very different than those expected for GMSB
SPS8 signal events, with larger differences observed for higher lifetime values.

\section{Statistical Analysis}
\label{sec:stats}

The photon pointing and time measurements are each sensitive to the possible
presence of photons from displaced decays of heavy, long-lived NLSP.
In addition, the measurements of \zdca \ and \tg \ are almost completely uncorrelated for prompt backgrounds.
The lack of correlation results from the fact that \zdca \ uses the spread of the
EM shower to precisely measure its centroids in the first two layers in the
EM calorimeter, while \tg \ uses the time reconstructed from the pulse-shape of
only the second-layer cell with the maximum energy deposit.
Using both variables to distinguish signal from background is therefore a powerful tool.

Since the \zdca \ distribution should be symmetric for both signal and background, the pointing distribution is folded
 by taking \abszdca as the variable of interest instead of \zdca .
The inputs to the statistical analysis are, therefore, the values of \abszdca \ and \tg \
measured for the photon selected in each event.

A full two-dimensional (2D) analysis of \abszdca \ versus \tg \ would require
populating a very large number of bins of the corresponding
2D space with both the background and signal models. Since the background
model is determined using data in control samples, which have limited
numbers of events, this approach is
impractical. Instead, the original 2D analysis is transformed into a
``N $\times$ 1D'' problem by using the \abszdca \ values to
define N mutually exclusive categories of photons, and then
simultaneously fitting the \tg \ spectra of each of the categories.
To optimize the sensitivity of the analysis, the categories
are chosen to divide the total sample of photons into
categories with different signal-to-background ratios.
 This approach is similar to that followed in
the ATLAS determination of the Higgs boson spin in the
\Hgg \ decay channel~\cite{Aad:2013xqa}.

An additional motivation for applying the ``N $\times$ 1D'' approach is to
simplify the task of modeling the overall background with an unknown
mixture of the background templates measured using the \Zee \ and \lowMET \ samples.
As shown in Fig.~\ref{fig:signaltiming}, these samples
used to model the various background contributions have
different \abszdca \ distributions, but very similar \tg \ distributions.
The minor \tg \ differences can be handled, as described in
Sec.~\ref{sec:systematics}, by including a small systematic uncertainty
on the \tg \ background shape. However, the \abszdca \ distribution of the total
background depends sensitively on the background composition. By
implementing the normalization of the background in each \abszdca \ category
as an independent, unconstrained nuisance parameter, the fitting
procedure eliminates the need
to predict the overall \ \abszdca \ distribution of the total
background, thereby avoiding the associated dependence on knowledge
of the background composition.

The binning in both \abszdca \ and \tg \
was chosen to optimize the expected sensitivity. It was found that using
six \abszdca \ categories and six \tg \ bins provides the analysis with good expected sensitivity,
without undue complexity. While the optimized choice of bin boundaries has almost no
dependence on $\Lambda$, there is some dependence on NLSP lifetime.
The analysis, therefore, uses two separate choices of binning, one for low lifetime
values ($\tau < 4$~ns) and one for high lifetime values ($\tau > 4$~ns).
The optimized category and bin boundaries for both cases are summarized
in Tables~\ref{tab:catedges} and \ref{tab:binedges}, respectively.

\begin{table*}[hbpt]
  \centering
 \caption{Values of the optimized ranges of the six \abszdca \ categories,
  for both low and high NLSP lifetime ($\tau$) values. }
  \begin{tabular}{ccccccc} \hline \hline
\ NLSP  \ & \multicolumn{6}{c}{Range of \abszdca \ values for each category [mm]}  \\ \cline{2-7}
\ Lifetime \ & \ \ \ \ Cat. 1 \ \ \ \ & \ \ \ \ Cat. 2 \ \ \ \ & \ \ \ \ Cat. 3 \ \ \ \ & \ \ \ \ Cat. 4 \ \ \ \ & \ \ \ \ Cat. 5 \ \ \ \ & \ \ \ \ Cat. 6 \ \ \ \  \\ \hline
$\tau < 4$~ns &  \ 0 -- 40 \ & \ 40 -- 80 \ & \ 80 -- 120 \ & \ 120 -- 160 \ & \ 160 -- 200 \ &  \ 200 -- 2000 \ \\
$\tau > 4$~ns & 0 -- 50 &  50 -- 100 & 100 -- 150 & 150 -- 200 &  200 -- 250 &  250 -- 2000 \\ \hline \hline
  \end{tabular}
  \label{tab:catedges}
\end{table*}

\begin{table*}[hbpt]
  \centering
 \caption{Values of the optimized ranges of the six \tg \ bins,
  for both low and high NLSP lifetime ($\tau$) values. }
  \begin{tabular}{ccccccc} \hline \hline
\ NLSP  \ & \multicolumn{6}{c}{Range of \tg \ values for each bin [ns]}  \\ \cline{2-7}
\ Lifetime \ & \ \ \ \ \ Bin 1 \ \ \ \ \ & \ \ \ \ \ Bin 2 \ \ \ \ \ & \ \ \ \ \ Bin 3 \ \ \ \ \ & \ \ \ \ \ Bin 4 \ \ \ \ \ & \ \ \ \ \ Bin 5 \ \ \ \ \ & \ \ \ \ \ Bin 6 \ \ \ \ \  \\ \hline
$\tau < 4$~ns &   $-$4.0 -- $+$0.5 &  0.5 -- 1.1 &  1.1 -- 1.3 &  1.3 -- 1.5 &  1.5 -- 1.8 &  1.8 -- 4.0 \\
$\tau > 4$~ns &  \ $-$4.0 -- $+$0.4 \ &  0.4 -- 1.2 &  1.2 -- 1.4 &  1.4 -- 1.6 &  1.6 -- 1.9 &  1.9 -- 4.0 \\ \hline \hline
  \end{tabular}
  \label{tab:binedges}
\end{table*}

The one-dimensional fits of the \tg \ distributions of the individual categories are performed simultaneously.
The signal normalization is represented by a single unconstrained signal-strength
parameter, $\mu$, that is correlated between all categories and defined as the fitted signal cross
section divided by the GMSB SPS8 prediction.  Thus, there are seven unconstrained parameters in
the fit, namely six separate nuisance parameters, one for each category, describing the background normalization,
and the signal strength $\mu$.

The analysis uses a likelihood model $\mathcal{L}(\mu,\theta)$ that is dependent on the signal
strength $\mu$ and the values of the nuisance parameters $\theta$. The model incorporates a
statistical Poisson component as well as Gaussian constraint terms for the nuisance parameters
associated with systematic uncertainties. The statistical model and procedure are implemented
within the HistFactory framework~\cite{Cranmer:1456844}. Two likelihood-based
test statistics $q_{0}$ and $q_{\mu}$ are calculated to find the $p_{0}$ values for the
background-only hypothesis and to set upper limits on the signal strength.

Asymptotic formulae based on Wilk's theorem are used to approximate the $q_{0}$ and $q_{\mu}$
distributions following the procedures documented in Ref.~\cite{Cowan:2010js}.
Tests of the background model's validity in the control regions and the signal region rely on the $p_{0}$ test statistic,
calculated from the observed $q_{0}$. In the absence of any excess, the $CL_{S}$ exclusions for each
signal type are calculated according to Ref.~\cite{Read:2002hq}.

To validate the statistical model and asymptotic forms of $q_{0}$ and $q_{\mu}$, unconditional
pseudo-experiment ensembles were generated from the background-only model and multiple
signal-plus-background models. Although the number of data events in the signal region is not large, deviations
from the asymptotic $\chi^{2}$ distribution of $q_{\mu}$ were shown to have a minimal impact
on the exclusion. The model accurately reconstructed the signal and background normalization
parameters and produced Gaussian distributions of the constrained nuisance parameters.

\section{Systematic Uncertainties}
\label{sec:systematics}

In the statistical analysis,  the background
normalization for each \abszdca \ category
is determined using an independent nuisance parameter.
Therefore, it is not necessary to include systematic uncertainties regarding
the normalization of the background, nor regarding its shape in the variable \abszdca.
As a result, the various systematic
uncertainties relevant for this analysis can be divided into two categories: so-called
``flat'' uncertainties are not a function of \abszdca \ and \tg \ and affect only the overall
signal yield, while ``shape'' uncertainties are those that are related to the shapes of the
unit-normalized \abszdca \ and \tg distributions for signal or to the shape
of the background \tg template.

\subsection{Signal yield systematic uncertainties}
\label{sec:sigyield}

The various flat systematic uncertainties affecting the signal yield are summarized in Table~\ref{tab:systematics}.
The uncertainty on the integrated luminosity is $\pm2.8$\% and is determined
with the methodology detailed in Ref.~\cite{Aad:2013ucp}.
The uncertainty due to the trigger is dominated by uncertainties on the dependence
on \abszdca \  of the efficiency of the hardware-based Level 1 (L1) trigger. The L1 calorimeter trigger~\cite{Achenbach:2008zzb}
uses analog sums of the channels grouped within projective trigger towers. This architecture leads to a small decrease
in L1 trigger efficiency for highly non-pointing photons, due to energy leakage from the relevant trigger towers.
The uncertainty on the impact of this dependence is conservatively set to the magnitude of the observed change in
efficiency in signal MC events versus \abszdca,  and  dominates the $\pm2$\% uncertainty on the trigger efficiency.

Following the  method outlined in Ref.~\cite{Aad:2013zba},
uncertainties on the signal efficiency, arising from the combined impact of uncertainties in the photon energy scale and resolution
and in the combined photon identification and isolation efficiencies, are determined to be $\pm1$\% and $\pm1.5$\%,
respectively.
An additional 4\% is included as a conservative estimate of the uncertainty in the identification efficiency
due to the non-pointing nature of the photons.
 This estimate  is derived from studies
of changes in the relevant variables measuring the shapes of the EM showers for non-pointing photons.
An uncertainty
on the signal yield of $\pm1.1$\%  results from
varying the  \met\ energy scale and resolution within their estimated uncertainties~\cite{Aad:2012re}.
The uncertainty on the signal efficiency due to MC statistics lies in
the range $\pm$(0.8--3.6)\% and the contribution due to the lifetime reweighting technique is in
the range $\pm$(0.5--5)\%, depending on the sample lifetime.

Variations in the calculated NLO signal cross sections times the signal acceptance and efficiency, at the level of
$\pm(9-14)$\% occur when varying the PDF set and factorization
and renormalization scales,  as described in Sec. ~\ref{sec:mc:simulation}. In the results, these uncertainties on the theoretical cross section
are shown separately, as hashed bands around the theory prediction.  Limits are quoted at the
points where the experimental results equal the value of the central theory prediction minus
one standard deviation of the theoretical uncertainty.

\begin{table}[hbpt]
  \centering
  \caption{Summary of relative systematic uncertainties that affect the normalization
  of the signal yield.  The last row summarizes the relative uncertainty on the theoretical
  cross section, and is treated separately, as explained in the text.}
  \begin{tabular}{lc} \hline \hline
  Source of uncertainty & Value  [\%]  \\ \hline
  Integrated luminosity & $\pm$ 2.8    \\
 Trigger efficiency   & $\pm$ 2  \\
 Photon \ET \ scale/resolution  &$\pm$ 1 \\
 Photon identification and isolation  & $\pm$ 1.5 \\
 Non-pointing photon identification  & $\pm$ 4 \\
 \MET \ reconstruction   & $\pm$ 1.1 \\
 Signal MC statistics   &\ \  \ \ \ \ $\pm$ (0.8--3.6) \ \ \ \ \ \   \\
 Signal reweighting   & $\pm$ (0.5--5) \\
 Signal PDF and scale uncertainties   & \ $\pm$ (9--14) \ \\    \hline  \hline
  \end{tabular}
  \label{tab:systematics}
\end{table}

\subsection{Signal shape systematic uncertainties}
\label{sec:sigshape}

The expected signal distributions are determined using
the GMSB SPS8 MC signal events. Therefore, limitations in
the MC simulation could lead to differences between data and MC events
in the predicted signal behavior. Any such discrepancies in the
shapes of the signal distributions must be handled by corresponding
systematic uncertainties on the signal shapes.
Since signal templates for both \abszdca \ and \tg \ are
used in the statistical analysis, systematic uncertainties on the signal shapes of both
must be taken into account in the fitting procedure.

The dominant systematic uncertainty on the shape of the signal \tg \ distribution arises from
the impact of the time reconstruction algorithm on
 the measurement of delayed signals. As discussed in Sec.~\ref{sec:delayed}, the use of fixed
OFC values causes a bias in the energy and time reconstructed
for signals that are sufficiently early or late compared
to the nominal time. For time shifts within $\pm 5$~ns
of the nominal time, the reconstructed energy decreases by less than 1\% and,
as a result, impacts on the measurements of the
photon energy and pointing are negligible. However, for time shifts of $\pm 5$~ns,
a bias in the time reconstruction of order 10\% of the shift is observed in the
analysis of satellite--satellite collisions.
Since the optimal filtering approach is equivalent
to a linearization of the optimization problem,
the expected form of the time bias is expected to be dominated
by the neglected quadratic terms in the Taylor expansion. Therefore, one expects
deviations in the time measurement to be small for small time shifts, over a
region where the linear approximation works well, and then to grow
roughly quadratically for larger time shifts. As a conservative estimate of
the systematic uncertainty on the time measurement due to these
effects, a linear dependence is assumed for the deviations, with an
amplitude of $\pm 10$\% of the reconstructed time. This uncertainty is
applied only to the signal time distribution, since the background time
shape is determined directly from data and therefore already includes
whatever impact is caused by the bias.

Another source of systematic uncertainty in the signal \abszdca \
and \tg \ shapes results from possible differences between the pileup conditions in data and signal MC events, even though
the MC signal samples are reweighted to
match the pileup distribution observed in the data.
The PV in GMSB SPS8 signal events should be correctly identified with high efficiency,
typically greater than 90\%,  due to the high \ET \ values
of the other SM particles produced in the SUSY decay chains. However,
the presence of pileup could still increase the
likelihood of incorrectly choosing the PV, potentially
impacting both the pointing and time measurements.
Nearby energy deposits that are not associated with the photon could also impact the
photon measurements, though these should be moderated by
the photon isolation requirements.  As a conservative estimate of the possible
influence of pileup, the signal shapes in the entire MC sample were compared with those in two roughly equally sized subsamples with differing levels of
pileup, chosen as those events with less than,  and those
with greater than or equal to, 13 reconstructed PV candidates. The small
differences observed are included as pileup-induced systematic uncertainties on the
signal template shapes.

To investigate the possible impact of the imperfect knowledge of the
material distribution in front of the calorimeter, one signal MC point
was simulated with the nominal detector description as well as
with a modified version that varies the material description within
the uncertainties. The signal distributions using the two detector
geometries are very similar, typically agreeing within a few percent. These variations are small compared to
the other systematic uncertainties on the signal shapes,
 and are therefore neglected.

Typical values of the total systematic uncertainties on the signal shapes are
around $\pm 10$\%, dominated by the impact of the time reconstruction algorithm on
 the measurement of delayed signals. These uncertainties have a very small impact
 on the overall sensitivity of the analysis, which is dominated by statistical
 uncertainties due to the limited size of the data sample in the signal region.

\subsection{Background shape systematic uncertainties}
\label{sec:bkgndshape}

The dominant uncertainty in the knowledge of the background template
shape arises from uncertainty in the background composition in the
signal region. As described in Sec.~\ref{sec:bgtemplates},  and
seen in Fig.~\ref{fig:signaltiming}, the EM shower development of electrons and photons
differs from that of jets and gives rise to somewhat different \tg shapes, and very different \abszdca shapes.  Therefore, the \tg \ and
\abszdca \ shapes for the total background depend on the background
composition.

The statistical analysis includes an independent normalization fit parameter
for the total background in each of the \abszdca \ categories. By this means,
the fit result avoids any dependence on the \abszdca \ distribution of the background
and it is not necessary to account for systematic uncertainties on the
background \abszdca shape. However, the background \tg \
shape is used in the fitting procedure, and therefore its associated systematic
uncertainties must be taken into account.

Since the time measurement is performed using only the second-layer cell of the
EM cluster with the maximum energy deposit,  it is expected that the time
should be rather insensitive to the details of the EM shower development and,
therefore, one would expect very similar time distributions for prompt electrons, photons and jets.
As seen in Fig.~\ref{fig:signaltiming}, this expectation
is largely satisfied since the \Zee \ and \lowMET \ \tg \ distributions are indeed very similar.
However, there are some effects that could cause a slight violation of
the assumption that the \tg \ distribution would be the same
for all prompt background sources.
Details of the EM shower development can indirectly impact the time measurement, for
example, due to cross-talk from neighboring cells. In addition, the time
measurement necessarily includes a correction for the time of flight from the PV;
therefore, misidentification of the PV can lead to shifts in the reconstructed time away
from the true time, and different background sources can have different rates
of PV misidentification. PV misidentification can also produce shifts in the pointing
measurement, introducing a non-zero correlation between \tg \ and \abszdca, even for
prompt backgrounds.

The \tg \ template from the diphoton sample with \lowMET \  includes contributions
from jets as well as EM objects and is taken as the nominal estimate of the
background \tg \ shape. The difference between this distribution and that of
the \Zee \ sample, which has a higher purity of EM objects, is taken as an estimate of the uncertainty due to the
background composition and is symmetrized to provide a symmetric
systematic uncertainty on the background \tg \ shape. The uncertainty is small for low time values, but reaches
almost $\pm 100$\% in the highest \tg \ bin. However, this uncertainty has little impact on the
overall sensitivity since the signal yield in the highest \tg \  bin is
much larger than the background expectation, even when this large background uncertainty is taken into account.

Another uncertainty in the background \tg \ shape arises from uncertainties in
the relative contributions of BB and BE events to the background in the signal region.
The definition of \tg \ for BB events as
the time of the photon with the maximum time value produces, as mentioned previously, a small
shift toward positive time values for such events, which does not exist for BE events.
Therefore, in constructing the total background \tg \ template, it is necessary to
appropriately weight the  \tg \ background templates measured separately for  BB and BE events in order to
match the background in the signal region. Since any signal can have a different
BB/BE composition than the background, the rate of BB and BE events in the signal region cannot simply be used
to determine the background composition.  However, the background-dominated  control regions can be used to
make an estimation of the background BB/BE composition. Comparing the various samples with $\MET < 75$~GeV,
BB events are estimated to contribute $(61\pm4)$\% of the total background in the signal region, where the uncertainty
conservatively covers the variations observed among various samples. Therefore, the nominal \tg \
background template is formed by appropriately weighting the BB and BE background distributions to
this fraction, with BB fractions varied by $\pm 4$\% to generate the $\pm 1\sigma$ variations on this shape
due to the uncertainty in the BB/BE background contributions. This systematic shape uncertainty reaches
less than $\pm 10$\% in the highest \tg \ bin and, therefore,  is much smaller than the dominant
uncertainty due to the background composition.

An additional systematic uncertainty on the background \tg \ shape arises from the event
kinematics.  As discussed in Sec.~\ref{sec:bgtemplates}, the minimum \ET \ requirements on the photons are
increased to 60~GeV for the \lowMET control sample, as opposed to 50~GeV for the signal region,
in order for the \lowMET \ control sample to select photons with kinematic properties more similar to the background photons expected in the signal region.
Systematic uncertainties on the \tg \  shape of the \lowMET \ sample
are determined by varying the photon \ET \ requirement up and down by
10~GeV. The three shapes agree quite well with each
other, with the observed variations reaching about $\pm40$\% in the highest time bin.

\section{Results and Interpretation}

Before examining the \abszdca \ and \tg \ distributions of the data in the signal region,
the two control regions, CR1 with $20 < \MET < 50$~GeV and CR2 with $50 < \MET < 75$~GeV,
are used to validate the analysis technique and background modeling.
Since the control regions should be dominated by background, their data distributions are expected to be well described by the background-only fit.

Table~\ref{tab:eventcounts} summarizes the number of selected events in CR1
and CR2, as well as those in the signal region (SR), showing that the control region
datasets are much larger than that of the signal region. It is of interest whether the
background modeling, including the assigned systematic uncertainties, is adequate to describe the
control region data within the statistical uncertainties of the data in the signal region.
Therefore, the fitting procedure was applied separately to the measured data
distributions in CR1 and CR2, scaled in each case to the total of 386 events of
the signal region.  The fit results for both control regions are in good agreement
with the background-only model for all tested signal points, validating the analysis methodology.

\begin{table}[hbpt]
  \centering
  \caption{Numbers of selected events in the two control
  regions (CR1 and CR2) and in the signal region (SR).}
  \begin{tabular}{ccc} \hline \hline
\ Sample \ & \ \ \MET \ range [GeV] \ \ & \ Number of  events \ \\ \hline
\ \ CR1 & $20 < \MET <  50$ & 	50751 \\
\ \ CR2 & $50 < \MET < 75$ & \ \ 3591 \\
SR & \ \ \ \  \ \  \ \ $\MET > 75$ & \ \ \ 386 \\ \hline \hline
  \end{tabular}
  \label{tab:eventcounts}
\end{table}

Figure~\ref{fig:SRdistributions} shows
the distributions of \zdca \ and \tg \ for the 386 events in the
signal region. %
The distributions of both variables are rather narrow, as expected for
background. In particular, there is no evidence for events in the tail of the \tg \ distribution
at positive times, as would be expected for a signal contribution due to delayed photons.
The \zdca \ distribution is quite symmetric around zero, as expected for both the signal
and for physics backgrounds. The
\abszdca \ and \tg \ distributions in the final, coarser binning are used as
inputs to the final fitting procedure and statistical analysis.

\begin{figure*}[hbpt]
\centering
\includegraphics[width=.48\textwidth]{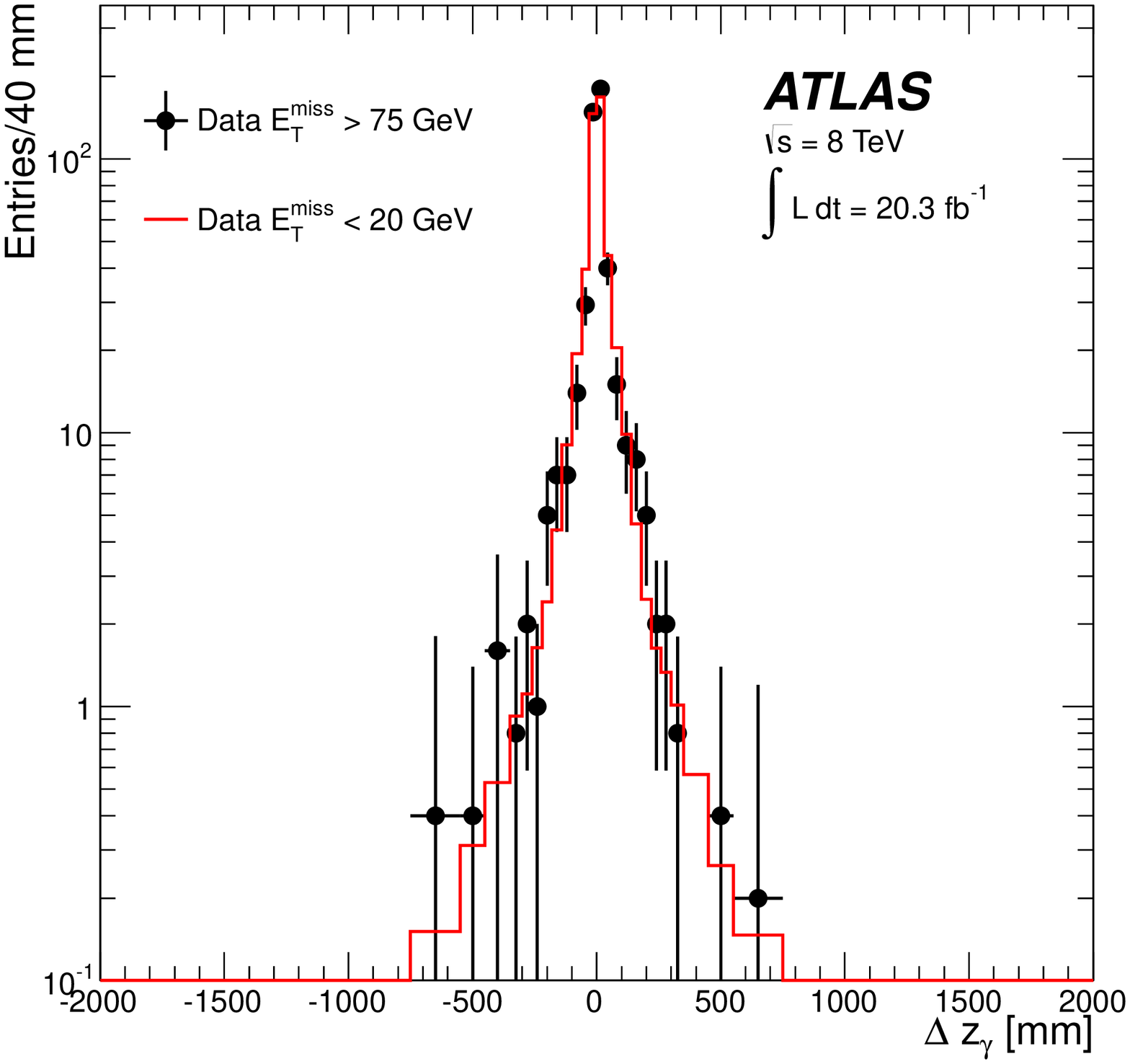}
\includegraphics[width=.48\textwidth]{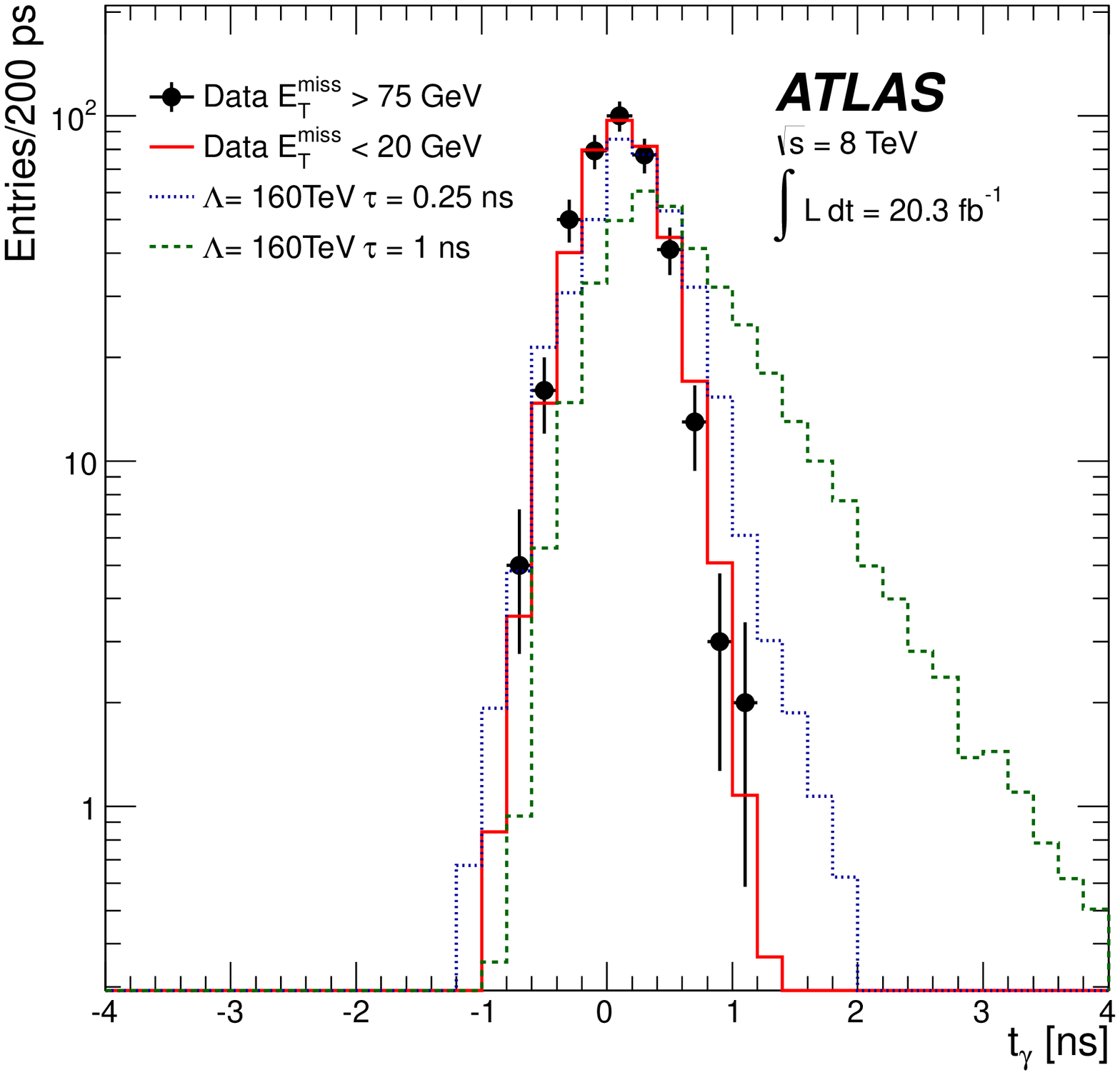}
\caption{Distributions of (left) \zdca \ and (right) \tg \ for the 386 events in the
signal region, defined with $\MET > 75$~GeV.  Superimposed are the data distributions for diphoton events with
$\MET < 20$~GeV, used to model the backgrounds, and the distributions for two example NLSP lifetime values in GMSB
SPS8 models with $\Lambda = 160$~TeV. The background and MC signal distributions are
scaled to the total number of data events in the signal region.}
\label{fig:SRdistributions}
\end{figure*}

Example results of fits to the signal region data are shown in Fig.~\ref{fig:SR_fit}, for the particular case of
$\Lambda = 100$~TeV and $\tau = 19$~ns. The figures show the results of the
signal-plus-background  (with $\mu = 1$) and background-only ($\mu = 0$) fits to the six \abszdca \ categories.
The signal-region data are in good agreement with the background-only model, and there is no evidence for a signal-like excess.

\begin{figure*}[hbpt]
\centering
\includegraphics[width=0.3\textwidth]{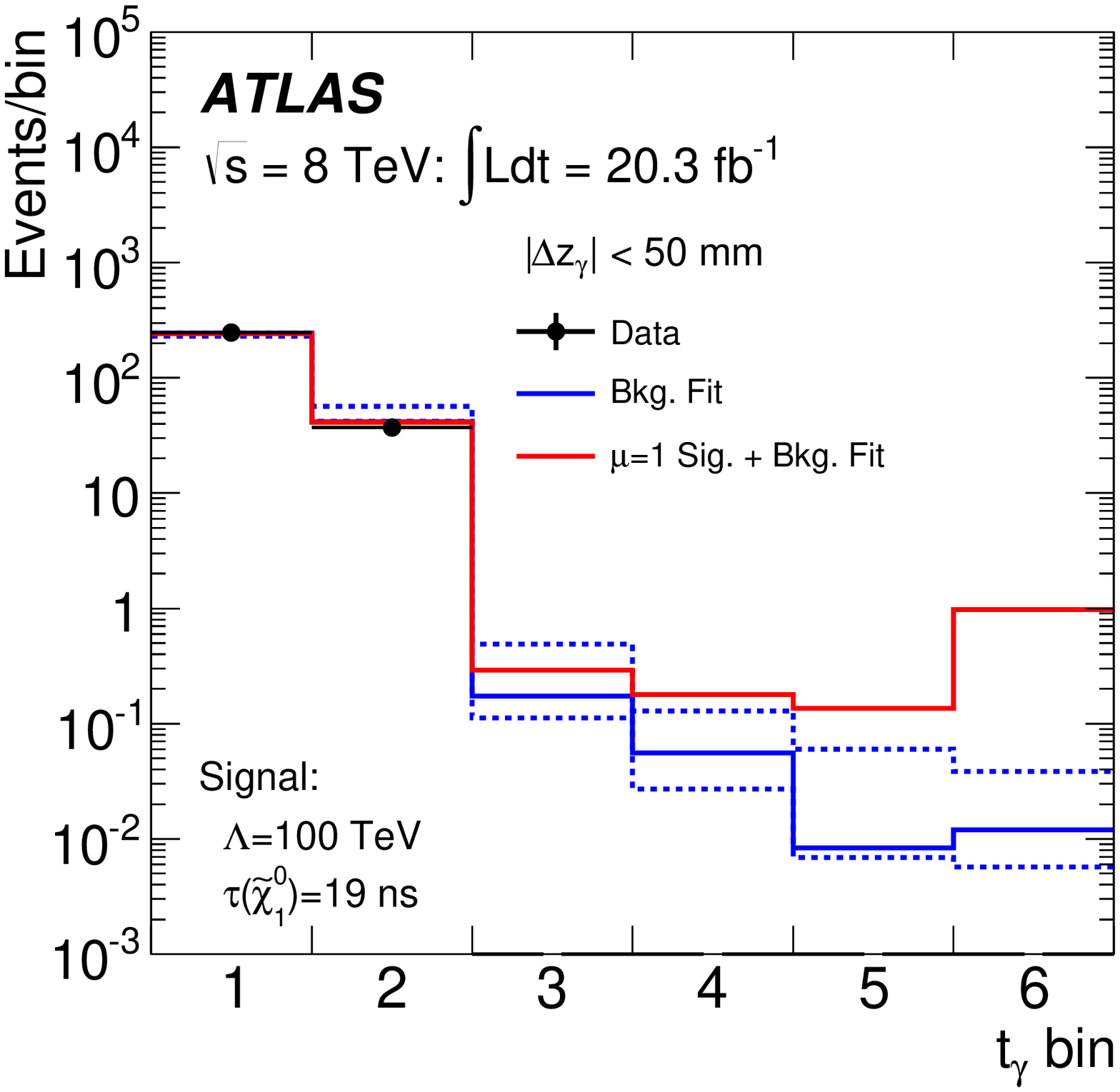}
\includegraphics[width=0.3\textwidth]{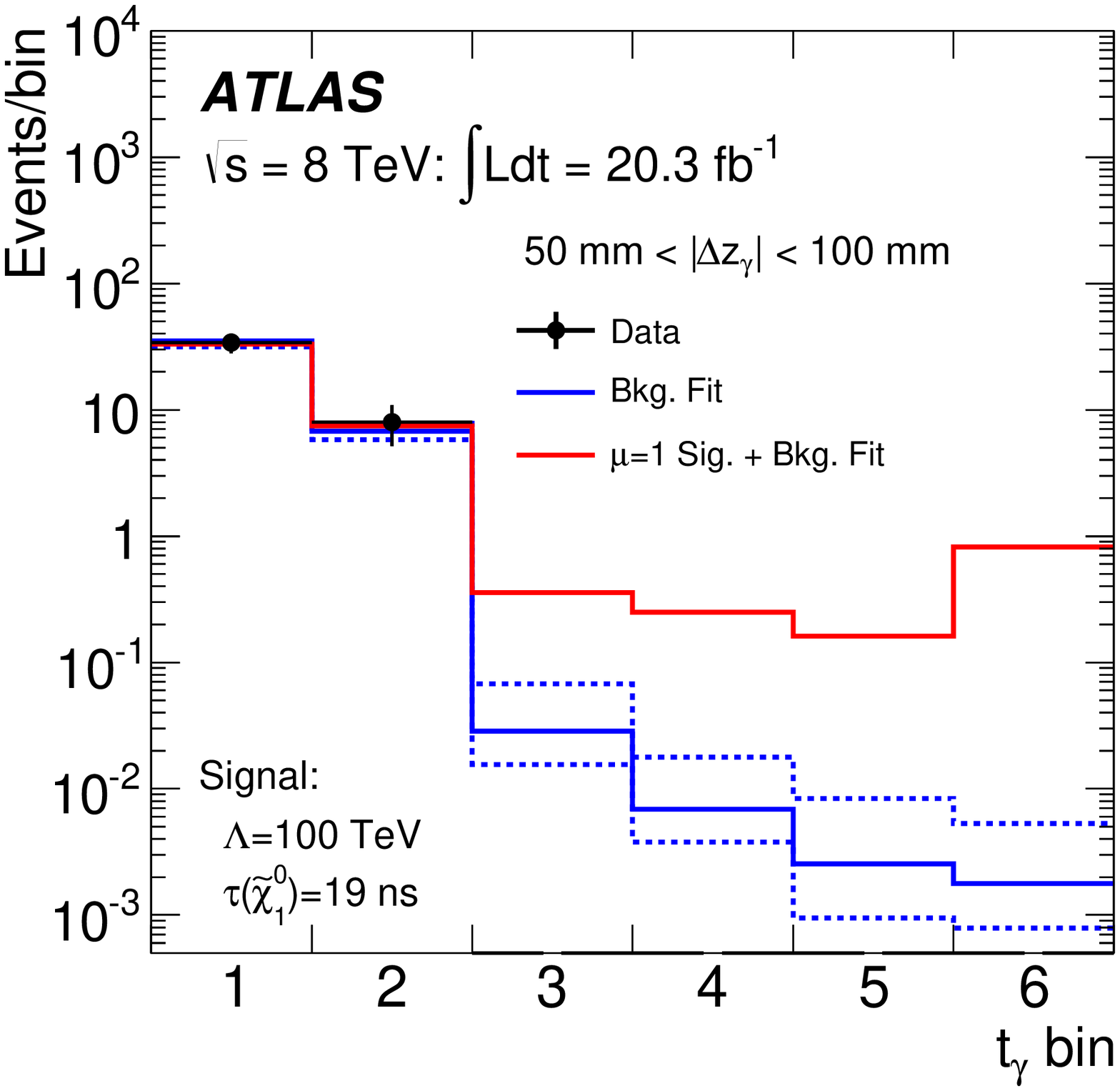}
\includegraphics[width=0.3\textwidth]{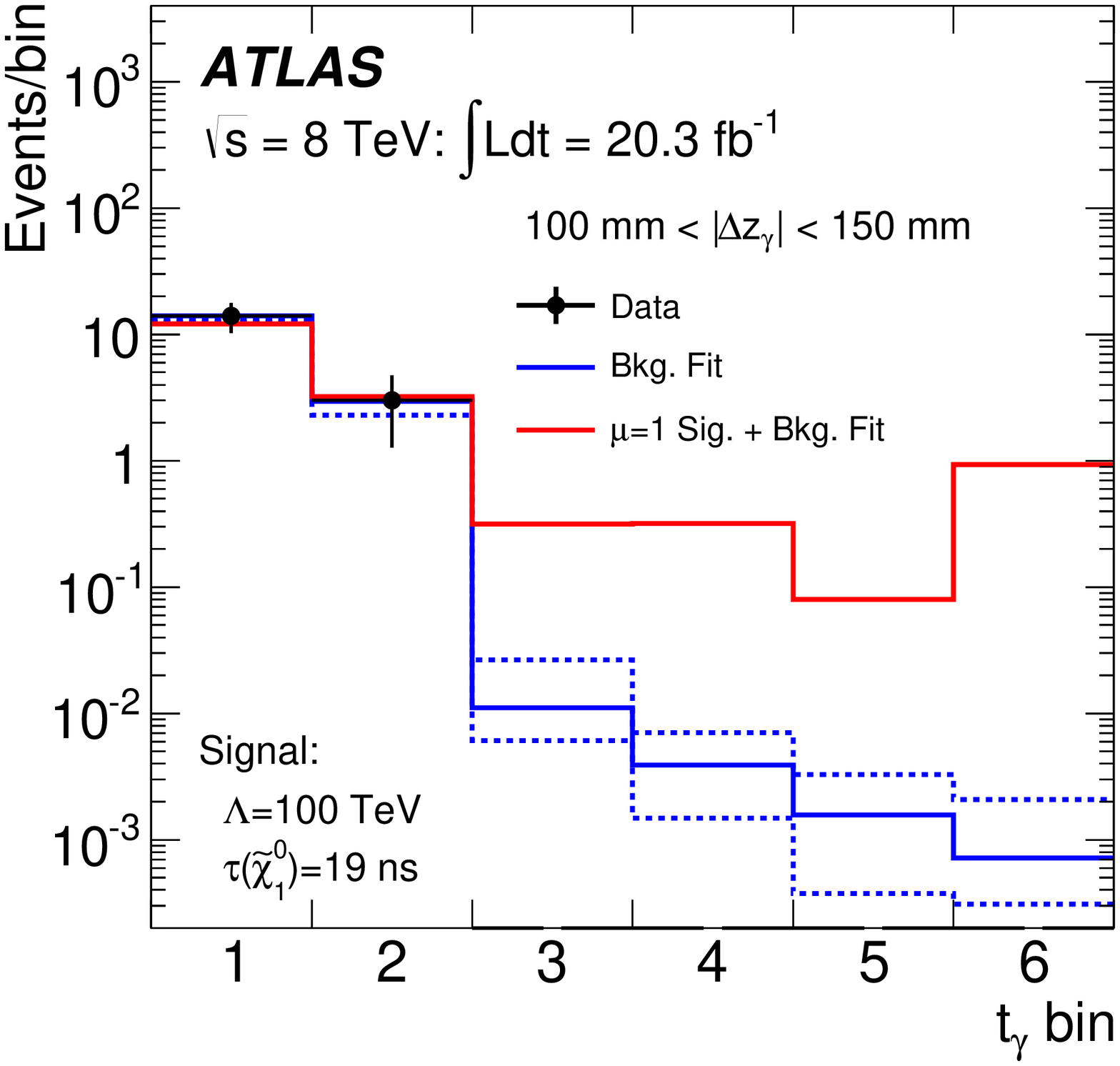} \\
\includegraphics[width=0.3\textwidth]{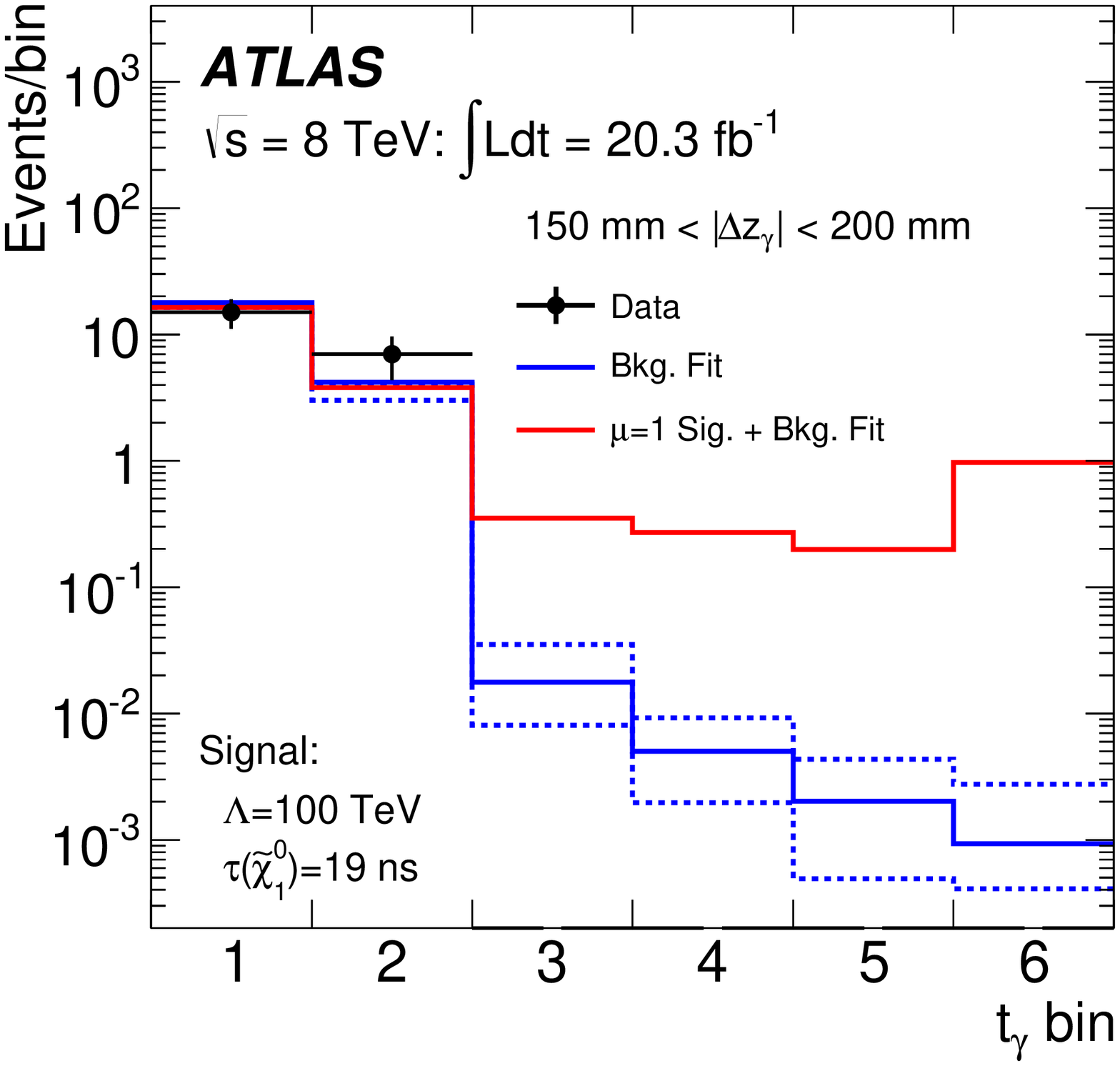}
\includegraphics[width=0.3\textwidth]{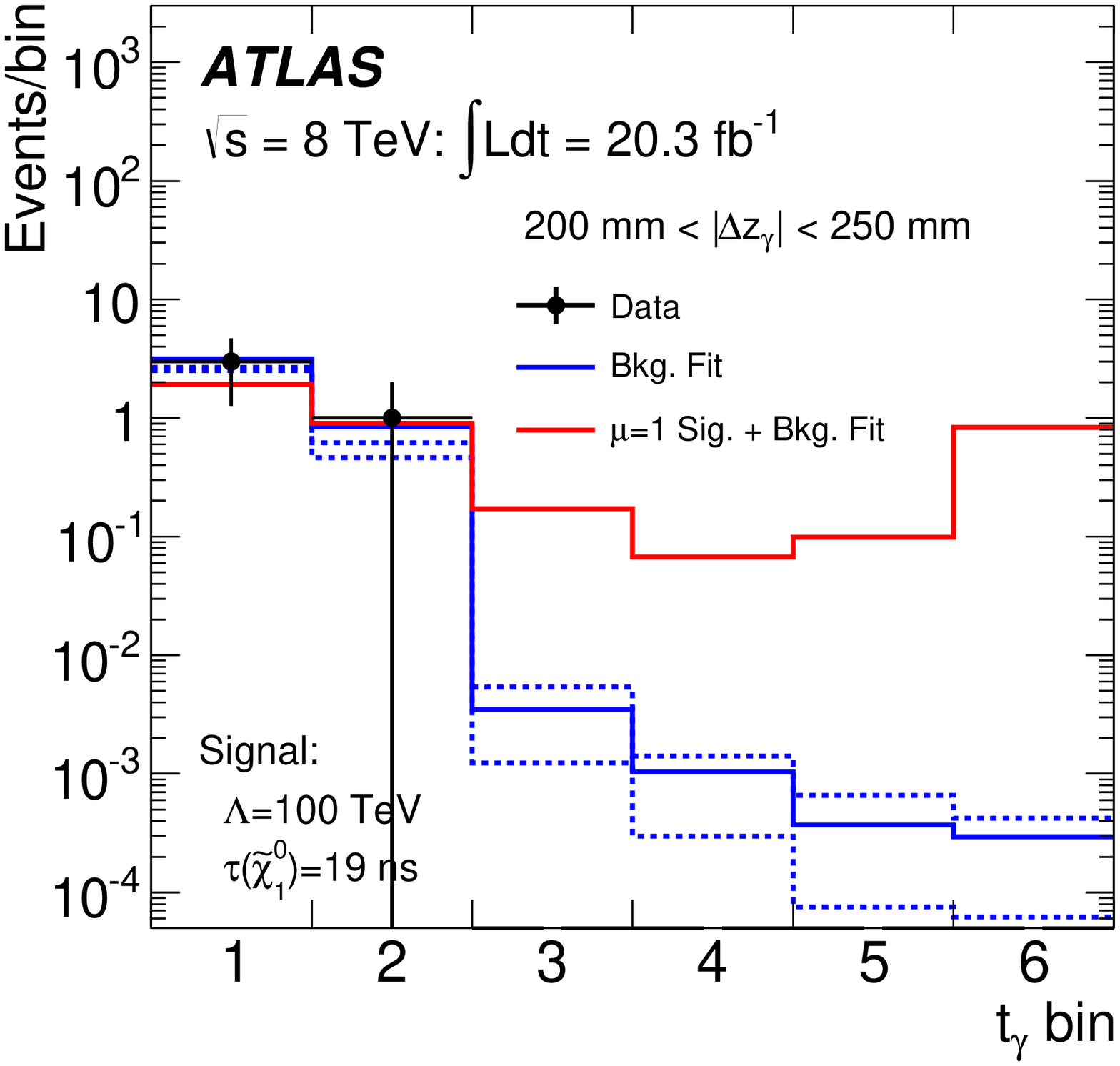}
\includegraphics[width=0.3\textwidth]{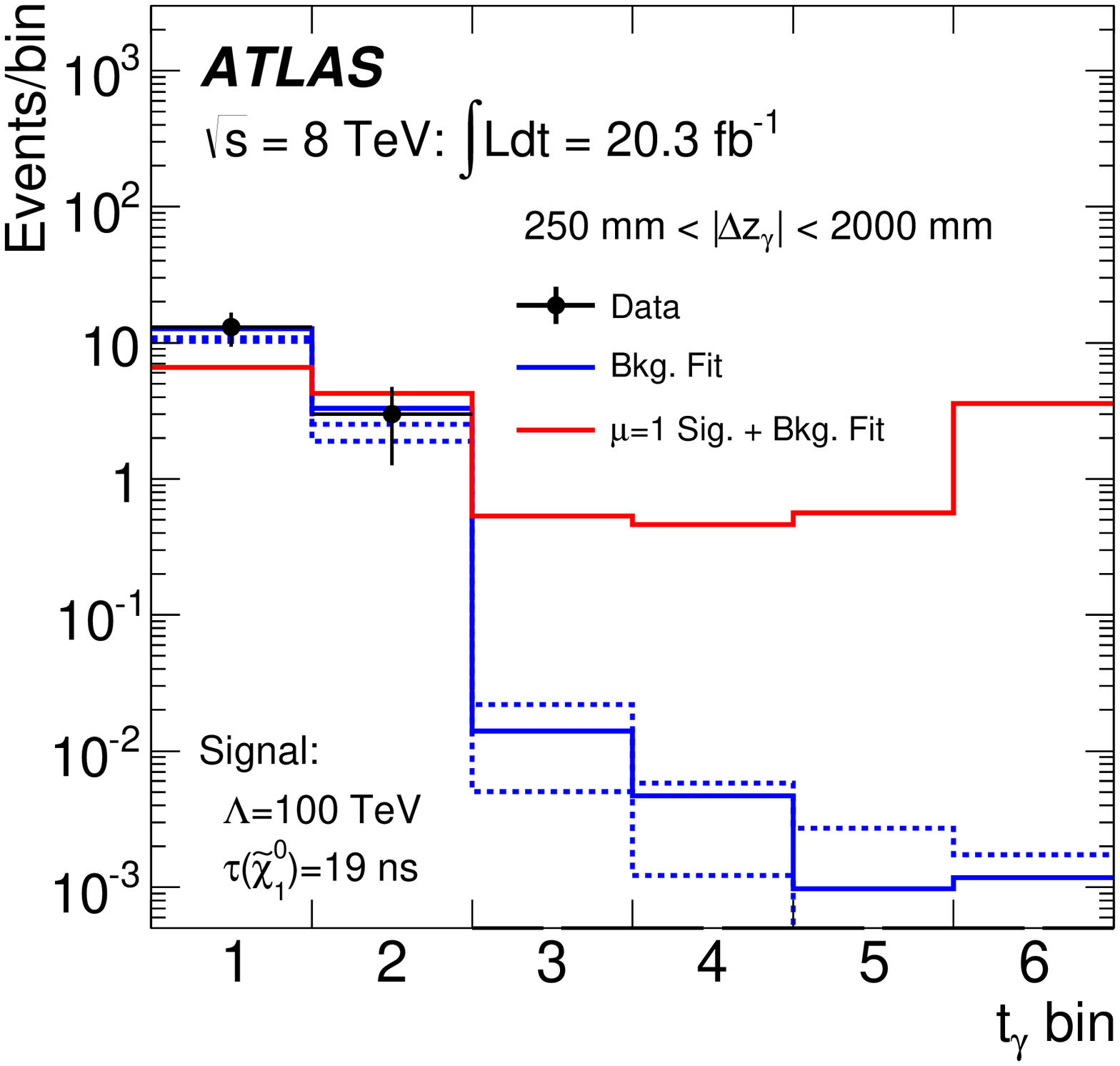}
\caption{Example fit to the signal-region data. The figures show the results of the signal-plus-background
fits with $\mu = 1$ to the six \abszdca \ categories, along with the background-only fit and
the $\pm1\sigma $ systematic shape variations (dashed lines). The signal model shown has
$\Lambda = 100$~TeV and $\tau = 19$~ns. The ranges of the
\tg \ bins are as defined in Table~\ref{tab:binedges}.}
\label{fig:SR_fit}
\end{figure*}

Fits to the data were performed for  $\tau$ values exceeding 250~ps, and for the range of relevant
$\Lambda$ values. The smallest $p_0$ value of 0.21, corresponding to an equivalent Gaussian
significance of $0.81\sigma$, was found for signal model parameters of $\Lambda = 100$~TeV and $\tau = 0.25$~ns.
Using ensembles of background-only pseudo-experiments, the probability of observing a $p_{0}$ value this small
or smaller from any one of the 640 signal points in the $\Lambda$--$\tau$ plane was calculated to be 88\%.

Figure~\ref{fig:SR_1Dlimits} shows, for  $\Lambda = 200$~TeV, the results of the signal-region fit
interpreted as 95\% confidence level (CL) limits on the number of signal events, as well as on the signal cross section,
as a function of \ninoone \ lifetime  (assuming the branching fraction $BR(\ninoone \ra \gamma \gravitino)=1$).
Each plot includes a curve indicating the GMSB SPS8 theory prediction for $\Lambda = 200$~TeV. The intersections
where the limits cross the theory prediction show that,  for $\Lambda = 200$~TeV, values of $\tau$ in the
range between approximately 0.3~ns and  20~ns are excluded at 95\% CL. The observed limits are in
good agreement with the expected limits, which are also shown in Fig.~\ref{fig:SR_1Dlimits}, along with
their $\pm 1\sigma$ and $\pm 2\sigma$ uncertainty bands. For large $\tau$ values, the 95\% CL limits
are close to the value of three events expected for a Poisson distribution with zero events observed,
indicating that
the results for high lifetimes are dominated by statistical uncertainties.  The limits on the number of signal
events are higher for low lifetimes, as it becomes more difficult to distinguish between the signal and background shapes.

\begin{figure*}[hbpt]
\centering
\includegraphics[width=0.48\textwidth]{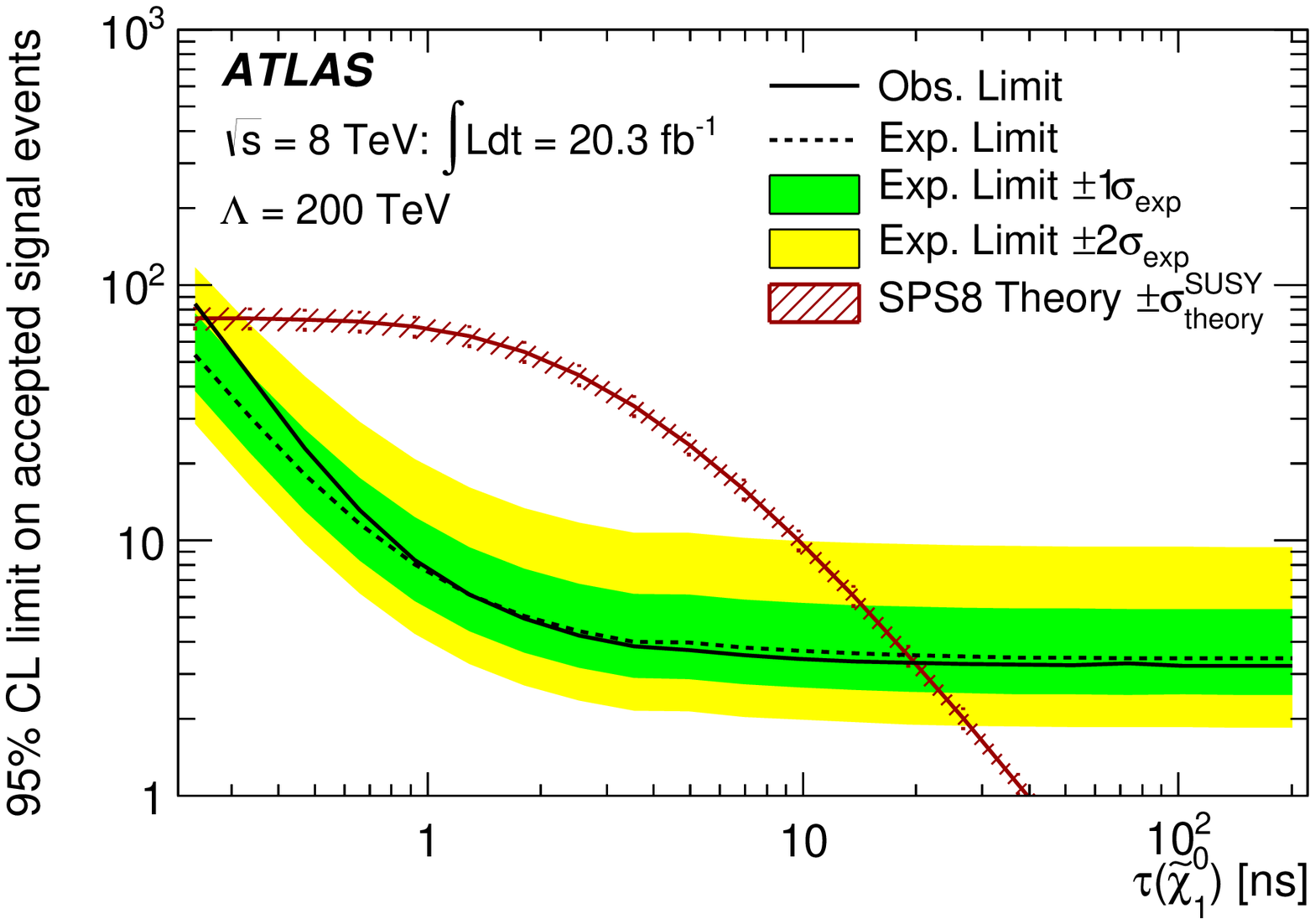}
\includegraphics[width=0.48\textwidth]{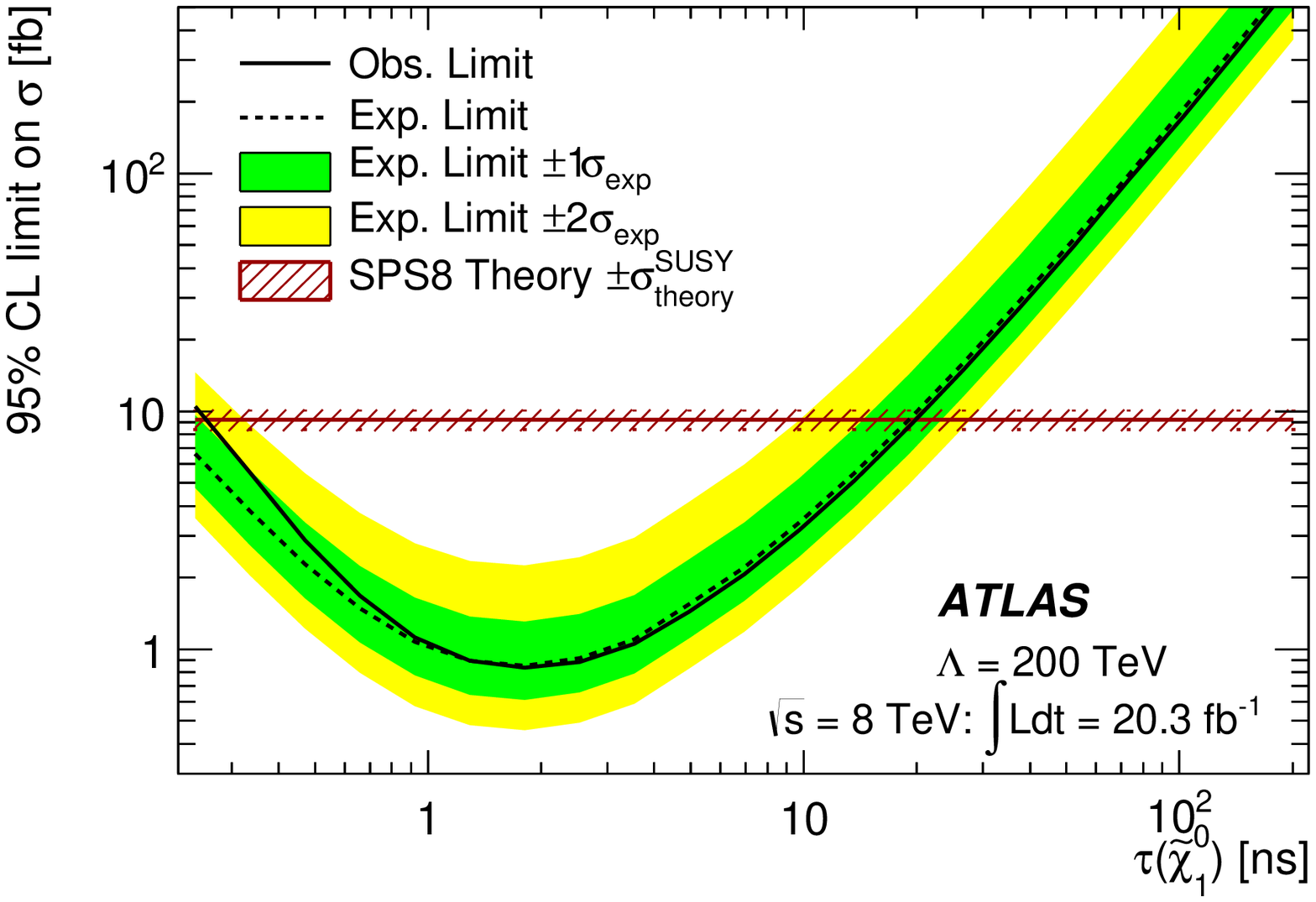}
\caption{The observed and expected 95\% CL limits on (left) the number of signal events and
(right) the GMSB SPS8 signal cross section, as a function of \ninoone \ lifetime, for the case of $\Lambda = 200$~TeV.
The regions above the limit curves are excluded at 95\% CL. The red bands show the GMSB
SPS8 theory prediction, including its theoretical uncertainty.}
\label{fig:SR_1Dlimits}
\end{figure*}

\begin{figure*}[hbpt]
\centering
\includegraphics[width=0.7\textwidth]{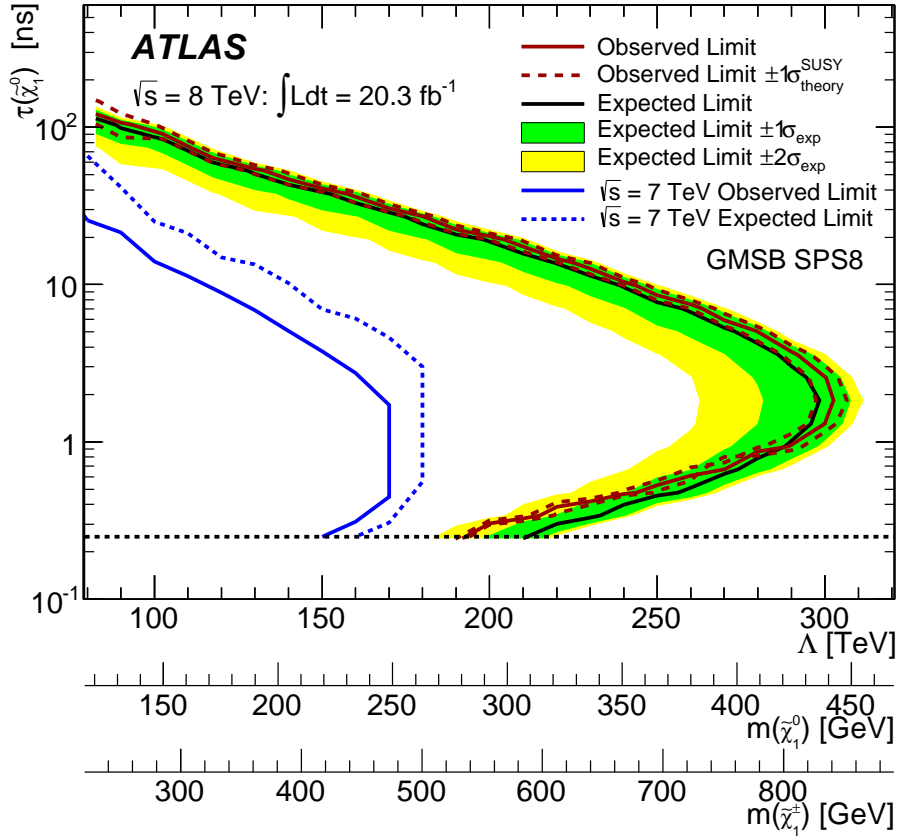}
\caption{The observed and expected 95\% CL limits in the two-dimensional GMSB signal space of
\ninoone \  lifetime versus $\Lambda$, the effective scale of SUSY breaking, and
also versus the corresponding \ninoone \ and \chinoonepm \ masses in the SPS8 model.
For comparison, the results from the 7~TeV analysis are also shown. The regions to the left of the
limit curves are excluded at 95\% CL. The horizontal dashed line indicates the lowest lifetime
value, namely $\tau = 250$~ps, for which the analysis is applied.}
\label{fig:SR_xsec_limit}
\end{figure*}

By repeating the statistical procedure for various $\Lambda$ and $\tau$
values, the limits are determined as a function of these GMSB SPS8 model parameters. The range of \ninoone \ lifetimes
tested is restricted to $\tau > 250$~ps to avoid the region of very low lifetimes where the shapes of the signal and background
distributions become very similar.
Figure~\ref{fig:SR_xsec_limit} shows the subsequent limits in the two-dimensional GMSB signal space of
\ninoone \  lifetime versus $\Lambda$, and
also versus the corresponding \ninoone \ and \chinoonepm \ masses in the GMSB SPS8 model. For example,
\ninoone \  lifetimes up to about 100~ns are excluded at 95\% CL for $\Lambda$ values in the range of about 80--100~TeV, as are
$\Lambda$ values up to about 300~TeV (corresponding to \ninoone \ and \chinoonepm \ masses of about 440~GeV and
840~GeV, respectively) for \ninoone \  lifetimes in the range of about 2--3~ns.
For comparison, the results from the ATLAS analysis of the 7~TeV dataset~\cite{Aad:2013oua}
are also shown in Fig.~\ref{fig:SR_xsec_limit}, indicating the significantly larger reach of the 8~TeV analysis.

\clearpage
\section{Conclusions}

A search has been made for evidence of non-pointing and delayed photons, such as would
arise in the decays of long-lived heavy neutral particles. The search, in the
$\gamma\gamma + \MET + X$ final state, uses the full data sample collected by ATLAS in 2012, corresponding
to an integrated luminosity of 20.3~\ifb \ of 8~TeV $pp$ collisions at the LHC.

The data are in good agreement with
the background-only fit and no evidence for non-pointing and delayed photons is observed.
The results are interpreted in the context of the GMSB SPS8 benchmark model,
in the plane of  $\tau$, the \ninoone \  lifetime, versus $\Lambda$, the effective scale of SUSY breaking, and
also versus the corresponding \ninoone \ and \chinoonepm \ masses.  The resultant 95\% CL exclusion limits
include values of $\tau$ in the range from 0.25~ns to about 100~ns, and values of $\Lambda$ in the range from
80~TeV to about 300~TeV.




\section{Acknowledgements}

We thank CERN for the very successful operation of the LHC, as well as the
support staff from our institutions without whom ATLAS could not be
operated efficiently.

We acknowledge the support of ANPCyT, Argentina; YerPhI, Armenia; ARC,
Australia; BMWFW and FWF, Austria; ANAS, Azerbaijan; SSTC, Belarus; CNPq and FAPESP,
Brazil; NSERC, NRC and CFI, Canada; CERN; CONICYT, Chile; CAS, MOST and NSFC,
China; COLCIENCIAS, Colombia; MSMT CR, MPO CR and VSC CR, Czech Republic;
DNRF, DNSRC and Lundbeck Foundation, Denmark; EPLANET, ERC and NSRF, European Union;
IN2P3-CNRS, CEA-DSM/IRFU, France; GNSF, Georgia; BMBF, DFG, HGF, MPG and AvH
Foundation, Germany; GSRT and NSRF, Greece; ISF, MINERVA, GIF, I-CORE and Benoziyo Center,
Israel; INFN, Italy; MEXT and JSPS, Japan; CNRST, Morocco; FOM and NWO,
Netherlands; BRF and RCN, Norway; MNiSW and NCN, Poland; GRICES and FCT, Portugal; MNE/IFA, Romania; MES of Russia and ROSATOM, Russian Federation; JINR; MSTD,
Serbia; MSSR, Slovakia; ARRS and MIZ\v{S}, Slovenia; DST/NRF, South Africa;
MINECO, Spain; SRC and Wallenberg Foundation, Sweden; SER, SNSF and Cantons of
Bern and Geneva, Switzerland; NSC, Taiwan; TAEK, Turkey; STFC, the Royal
Society and Leverhulme Trust, United Kingdom; DOE and NSF, United States of
America.

The crucial computing support from all WLCG partners is acknowledged
gratefully, in particular from CERN and the ATLAS Tier-1 facilities at
TRIUMF (Canada), NDGF (Denmark, Norway, Sweden), CC-IN2P3 (France),
KIT/GridKA (Germany), INFN-CNAF (Italy), NL-T1 (Netherlands), PIC (Spain),
ASGC (Taiwan), RAL (UK) and BNL (USA) and in the Tier-2 facilities
worldwide.

\addcontentsline{toc}{part}{Bibliography}
\bibliographystyle{atlasBibStyleWoTitle.bst}

\providecommand{\href}[2]{#2}\begingroup\raggedright\endgroup

\onecolumngrid
\begin{flushleft}
{\Large The ATLAS Collaboration}

\bigskip

G.~Aad$^{\rm 85}$,
B.~Abbott$^{\rm 113}$,
J.~Abdallah$^{\rm 152}$,
S.~Abdel~Khalek$^{\rm 117}$,
O.~Abdinov$^{\rm 11}$,
R.~Aben$^{\rm 107}$,
B.~Abi$^{\rm 114}$,
M.~Abolins$^{\rm 90}$,
O.S.~AbouZeid$^{\rm 159}$,
H.~Abramowicz$^{\rm 154}$
% ...
\bigskip
\\
% ...
$^{11}$ Institute of Physics, Azerbaijan Academy of Sciences, Baku, Azerbaijan\\
% ...
$^{85}$ CPPM, Aix-Marseille Universit{\'e} and CNRS/IN2P3, Marseille, France\\
% ...
$^{90}$ Department of Physics and Astronomy, Michigan State University, East Lansing MI, United States of America\\
% ...
$^{107}$ Nikhef National Institute for Subatomic Physics and University of Amsterdam, Amsterdam, Netherlands\\
% ...
$^{113}$ Homer L. Dodge Department of Physics and Astronomy, University of Oklahoma, Norman OK, United States of America\\
$^{114}$ Department of Physics, Oklahoma State University, Stillwater OK, United States of America\\
% ...
$^{117}$ LAL, Universit{\'e} Paris-Sud and CNRS/IN2P3, Orsay, France\\
% ...
$^{152}$ Institute of Physics, Academia Sinica, Taipei, Taiwan\\
% ...
$^{154}$ Raymond and Beverly Sackler School of Physics and Astronomy, Tel Aviv University, Tel Aviv, Israel\\
% ...
$^{159}$ Department of Physics, University of Toronto, Toronto ON, Canada\\
% ...
\end{flushleft}


\end{document}